\documentclass[a4paper,fleqn,usenatbib]{mnras}

\usepackage{newtxtext}

\usepackage[T1]{fontenc}
\usepackage{ae,aecompl}

\usepackage{graphicx}	
\usepackage{amsmath}	
\usepackage{amssymb}	

\def\spose#1{\hbox to 0pt{#1\hss}}
\def\lta{\mathrel{\spose{\lower 3pt\hbox{$\mathchar"218$}}
     \raise 2.0pt\hbox{$\mathchar"13C$}}}
\def\gta{\mathrel{\spose{\lower 3pt\hbox{$\mathchar"218$}}
     \raise 2.0pt\hbox{$\mathchar"13E$}}}

\def\p0{\phantom{0}}

\title[Chromospheric Activity in 55~Cancri~I.]
{Chromospheric Activity in 55~Cancri: \\
I.~Results from Theoretical Wave Studies}

\author[Fawzy and Cuntz]{
Diaa E. Fawzy$^{1}$ and Manfred Cuntz$^{2}$\thanks{E-mail: cuntz@uta.edu (MC)}
\\
$^{1}$Faculty of Engineering, Izmir University of Economics, 35330 Izmir, Turkey \\
$^{2}$Department of Physics, University of Texas at Arlington, Arlington, TX 76019, USA
}

\date{Accepted XXX. Received YYY; in original form ZZZ}

\pubyear{2020}

\begin{document}
\label{firstpage}
\pagerange{\pageref{firstpage}--\pageref{lastpage}}
\maketitle

  \begin{abstract}
We present theoretical models of chromospheric heating for 55~Cancri, an orange dwarf
of relatively low activity.  Self-consistent, nonlinear and time-dependent {\it ab-initio}
numerical computations are pursued encompassing the generation, propagation, and
dissipation of waves.  We consider longitudinal waves operating among arrays of flux tubes
as well as acoustic waves pertaining to nonmagnetic stellar regions.  Additionally, flux
enhancements for the longitudinal waves are also taken into account as supplied by transverse
tube waves.  The Ca~II~K fluxes are computed (multi-ray treatment) assuming partial redistribution
as well as time-dependent ionization.  The self-consistent treatment of time-dependent ionization
(especially for hydrogen) greatly impacts the atmospheric temperatures and electron densities
(especially behind the shocks); it also affects the emergent Ca~II fluxes.  Particularly, we focus
on the influence of magnetic heating on the stellar atmospheric structure and the emergent Ca~II
emission, as well as the impact of nonlinearities.  Our study shows that a higher photospheric
magnetic filling factor entails a larger Ca~II emission; however, an increased initial wave energy flux
(e.g., associated with mode coupling) is of little difference.  Comparisons of our theoretical results
with observations will be conveyed in forthcoming Paper~II.
  \end{abstract}




\section{Introduction}

In the previous decades a large number of studies has been presented indicating
the significance of wave heating in outer stellar atmospheres for stars of different
activity levels, including relatively inactive stars
\citep[e.g.,][including references therein and subsequent work]{nar96}; see also
more recent observations by \cite{fre14} and \cite{kay18}.
The general picture that emerged indicates that the atmospheres of high-activity stars
are dominantly heated by magnetic processes (including magnetic waves), whereas for
the atmospheric heating of low-activity stars non-magnetic processes may play
a more prominent role \citep[e.g.,][]{sch95,buc98,ramc03},
although the role of magnetic phenomena in those stars
is expected to be significant as well \citep[e.g.,][]{jud93,jud98}.
When stars age, the relative importance of atmospheric magnetic processes
tends to subside, a process closely related to the evolution of angular momentum
\citep[e.g.,][]{mar86,har87,kep95,cha97,wol97,mit18}.
Consequently, main-sequence stars of advanced age, as
e.g. old G and K stars, are targets of great interest for the study of the
relative importance of magnetic and non-magnetic wave atmospheric heating.

As pointed out by, e.g., \cite{lin83}, \cite{sch00}, as well as a large array of
other work, magnetic heating in stellar atmospheres is strongly correlated to
increased outer atmospheric emission.  Increased stellar rotation usually entails
an increased photospheric magnetic flux (${\propto}{B_0}{f_0}$), where $B_0$ is
the photospheric magnetic field strength and $f_0$ is the magnetic filling factor.
Therefore, in a statistical sense, it is possible to link ${B_0}{f_0}$ to both the stellar rotation
period $P_{\rm rot}$ and age \citep[e.g.,][]{noy84,mar89,mon93,saa96},
on the one hand, and to the emergent chromospheric emission flux
\citep[e.g.,][]{sch89,jor97,faw02b}, on the other hand.

In this work, we pursue studies for 55~Cancri (55~Cnc, $\rho^1$~Cnc), a G8~V star \citep{gon98}.
55~Cnc is an old main-sequence star with a mass and luminosity considerably smaller than the Sun
and a slow rotation rate (see below); this star also meets the characteristics of an orange dwarf.
In this work, we study the relative importance of magnetic heating and non-magnetic
heating (i.e., acoustic waves) in consideration of 55~Cnc's low activity level, including
the emergence of chromospheric emission.
Our paper is structured as follows: In Section~2, we convey our theoretical approach,
including a description of the acoustic and magnetic wave energy generation, the flux
tube models, and the computation of the emergent Ca~II emission fluxes in response
to our wave calculations.  In Section~3, we present our results and discussion, including
relevant aspects of time-averaged atmospheres.  In Section~4, we give our
summary and conclusions.


\section{Theoretical Approach}


\subsection{Stellar Parameters}

In this work, we pursue theoretical atmospheric studies of 55~Cnc,
a G8~V star \citep{gon98}; see Table~1 for details.  A recent study by \cite{bra11} conveys
a mass of 0.91~$M_\odot$; hence, 55~Cnc can be classified as an orange dwarf; those stars
characterized by a mass range of about $0.80 \pm 0.13$ \citep{gra05}.  Furthermore, 55~Cnc
is considerably older than the Sun.  Previous estimates of 55~Cnc's age are
8.6~Gyr \citep{mam08} and 10.2~Gyr \citep{bra11}; see \cite{bou18} for discussion.

Previous work about the stellar rotation period indicates values of 42.2~d \citep{hen00}
and 38.8~d \citep{bou18}; clearly, these values of slow rotation are closely connected to
55~Cnc's advanced age, see, e.g., \cite{sku72} and subsequent work.  In the framework of
our models, the stellar rotation period as previously derived also allows default estimates
of the stellar photospheric magnetic filling factor (see Sect. 2.3).

%
\begin{table}
	\centering
	\caption{Stellar Parameters}
\begin{tabular}{lcl}
\noalign{\smallskip}
\hline
\noalign{\smallskip}
Parameter  &  Value  &  Reference  \\
\noalign{\smallskip}
\hline
\noalign{\smallskip}
Spectral Type              &   G8~V  &  \cite{gon98}  \\
Effective Temperature~(K)  &   5165  &  \cite{lig16}  \\
...                        &   5172  &  \cite{yee17}  \\
Color~$(B-V)$              &   0.87  &  \cite{hof95}  \\
Mass~($M_\odot$)           &   0.91  &  \cite{bra11}  \\
Surface Gravity~(log cgs)  &   4.37  &  \cite{gra03}  \\
...                        &   4.43  &  \cite{yee17}  \\
Metallicity~[Fe/H]         &   0.35  &  \cite{yee17}  \\
Age~(Gyr)                  &   8.6   &  \cite{mam08}  \\
...                        &  10.2   &  \cite{bra11}  \\
Rotation Period~(d)        &  42.2   &  \cite{hen00}  \\ 
...                        &  38.8   &  \cite{bou18}  \\             
\noalign{\smallskip}
\hline
\noalign{\smallskip}
\multicolumn{3}{p{0.95\columnwidth}}{
Note:
See references for information on the uncertainty bars and
background information on the adopted methodology.  No claim
is made about the completeness of this list.
}
\end{tabular}
\end{table}

%
\begin{table}
	\centering
	\caption{Summary of Acronyms}
\begin{tabular}{ll}
\noalign{\smallskip}
\hline
\noalign{\smallskip}
Acronym  &  Definition  \\
\noalign{\smallskip}
\hline
\noalign{\smallskip}
ACW   &  Acoustic Wave                        \\
LTE   &  Local Thermodynamic Equilibrium      \\
LTW   &  Longitudinal Flux Tube Wave          \\
MFF   &  Magnetic Filling Factor ($f_0$)      \\
NLTE  &  Non-Local Thermodynamic Equilibrium  \\
NTDI  &  Non-Time-Dependent Ionization        \\
PRD   &  Partial Redistribution               \\
TDI   &  Time-Dependent Ionization            \\
TTW   &  Torsional Flux Tube Wave             \\
\noalign{\smallskip}
\hline
\end{tabular}
\end{table}


\subsection{Acoustic and Magnetic Wave Energy Generation}

A crucial aspect of this study pertains to the calculation of the initial acoustic
and magnetic wave energy fluxes; notably, model simulations for the generation of
longitudinal flux tube waves (LTWs)\footnote{See Table~2 for a summary of the
acronyms.}.  Regarding acoustic waves (ACWs), previous
calculations for wave energy fluxes have been given by \cite{mus94} and
\cite{ulm96}.  The latter authors computed both acoustic frequency spectra
and total acoustic fluxes on the basis of mixing-length convection zone models
for a large range of stars, including main-sequence stars.  The results were
identified to depend on the stellar effective temperature, surface gravity,
metallicity and the mixing-length parameter $\alpha_{\rm ML}$.

Following subsequent work by \cite{ste09a,ste09b}, who employed detailed 3-D
convective models, the value of $\alpha_{\rm ML} = 1.8$ has become increasingly
established; it has also been previously used by \cite{faw12}.  For 55~Cnc
this approach yields an acoustic wave energy flux of
$F_{\rm AC} = 3.3 \times 10^7$ erg~cm$^{-2}$~s$^{-1}$ with
a representative wave period of $P_{\rm AC} = 60$~s.  This latter value
has also been adopted for our monochromatic wave simulations intended for
comparison with the more detailed models based on wave spectra;
see Fig.~1 for results on the acoustic wave energy spectra.

Akin to our previous work, we calculate wave energy fluxes and spectra
for LTWs based on previous work by \cite{mus95}, \cite{ulm98}, and \cite{ulm01}, among others.
In this approach, stellar turbulence is described via an extended Kolmogorov
turbulent energy spectrum and the modified Gaussian frequency factor; see
\cite{mus95} for a detailed discussion of the potential and limitations of
this approach.  According to that work, the interaction between the magnetic flux
tubes and the surrounding turbulent medium results in the generation of LTWs,
among other types of waves.

At the start of the procedure, initial stellar
magnetic flux tube models are constructed (see Sect. 2.3) with the external
pressure fluctuations considered to be responsible for the squeezing of the
flux tube; they are represented by a superposition of a sufficiently high number
of partial waves.  Again, an extended Kolmogorov spatial turbulent energy spectrum
and the modified Gaussian frequency factor is applied.  Using the stellar
parameters of 55~Cnc as input (see Sect. 2.1), we find
$F_{\rm LTW} = 1.7 \times 10^8$ erg~cm$^{-2}$~s$^{-1}$ (upward directed flux)
for a magnetic field strength of $B_0$ = 1698~G (see Sect. 2.3), independent of the value
of the filling factor $f_0$, with a representative wave period of
$P_{\rm LTW} = 60$~s as also obtained
for acoustic waves.  The wave frequency spectrum for LTWs is given in Fig.~1.

%
\begin{figure*}
  \includegraphics[width=0.50\linewidth]{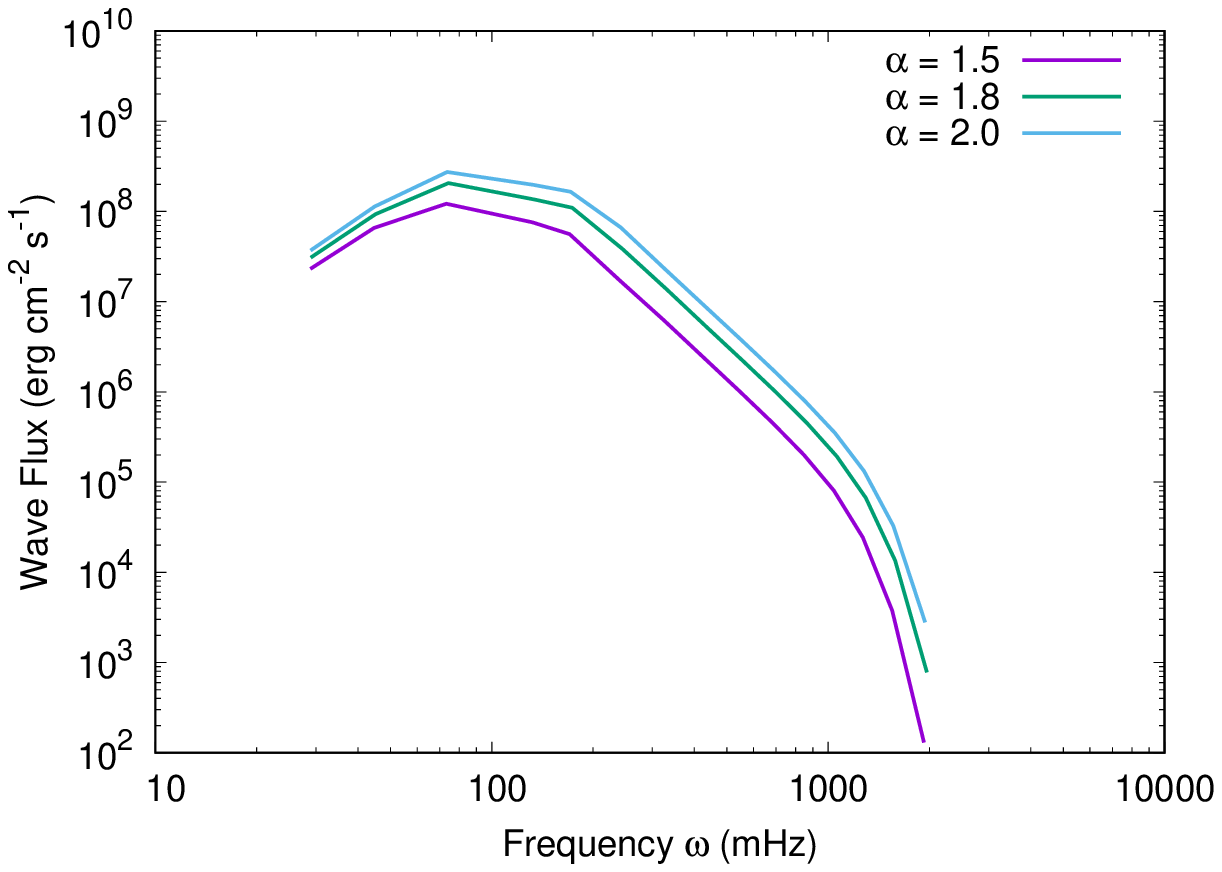}  \\
  \includegraphics[width=0.50\linewidth]{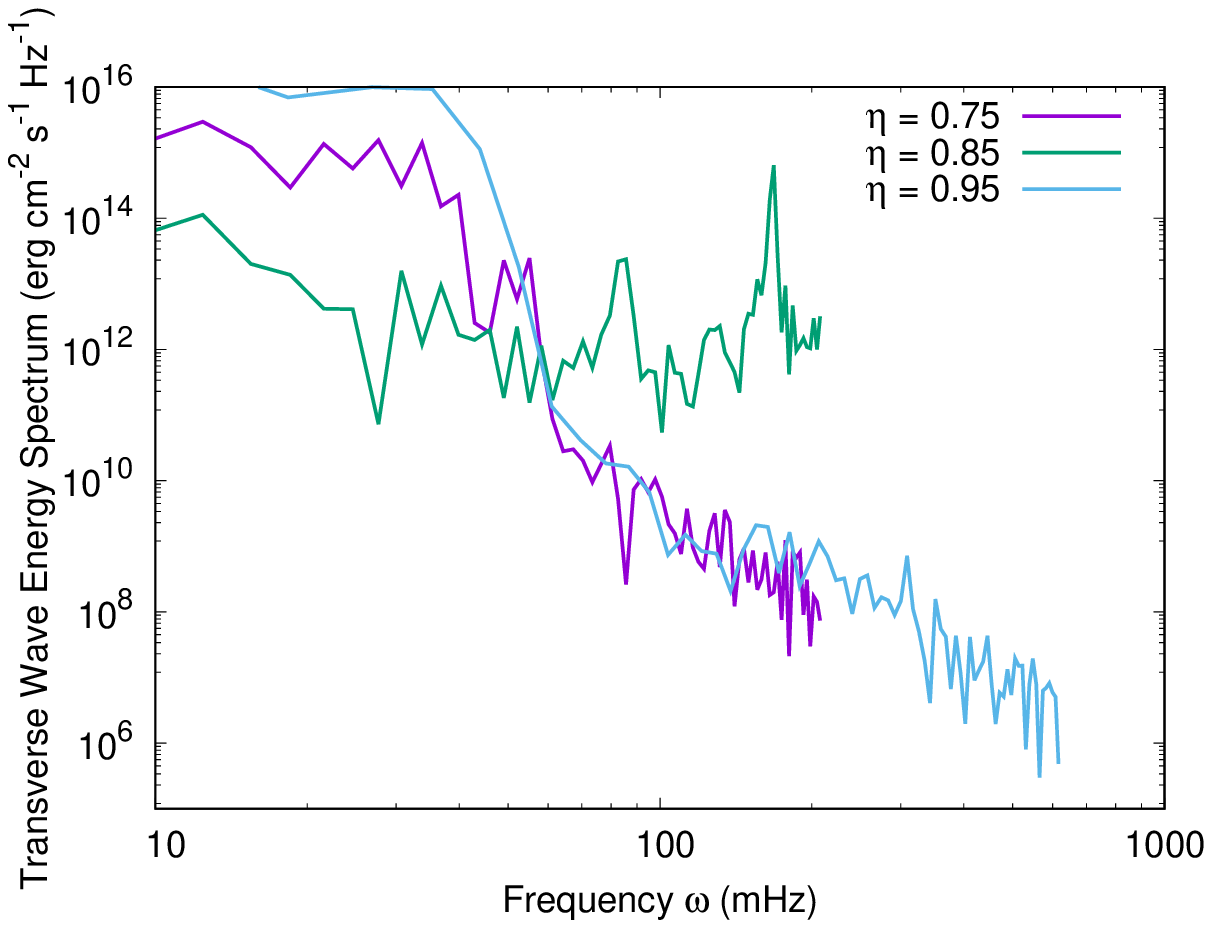}
  \caption{Initial wave energy spectra for acoustic wave models (top) and
longitudinal flux tube models (bottom).}
\end{figure*}


\subsection{Flux Tube Models}

The nonlinear interactions between the magnetic structures given as flux tubes
on the stellar surface and the surrounding turbulence entail the generation of waves
that are able to heat the stellar atmospheres, see, e.g., \cite{ulm01} and related studies.
This approach has also been implemented for the current computations.  The computations are started by
constructing thin magnetic flux tubes that are vertically oriented in a magnetic-free surrounding.
Therefore, we need to specify four basic parameters; namely, the stellar effective temperature $T_{\rm eff}$,
the surface gravity $g$, the magnetic field strength at the stellar surface B$_{0}(z=0~{\rm km})$, and the
surface magnetic filling factor $f_{0}(z=0~{\rm km})$.  Regarding our models, the latter determines the
geometry of the magnetic flux tubes as well as the tube spreading at the upper layers.  It is noteworthy
that the efficiency of wave heating of the upper layers critically depends on the flux tube geometry
\citep[e.g.,][]{faw98}.

The value of the magnetic field strength, B$_{0}(z=0~{\rm km})$, is of critical importance for the
generation of the magnetic wave energy as the efficiency decreases for rigid tubes of higher field strengths.
The determination of the value $B_{0}(z=0~{\rm km})$ from observations is not straightforward; however,
the value of $B_{0}f_{0}$ can be determined with a relatively high accuracy\footnote{Main-sequence stars of
different activity levels but of the same spectral type are expected to exhibit (approximately) the same value
of $B_0$, as set by the stellar photospheres.  Yet increased activity levels typically correspond to larger photospheric
magnetic filling factors $f_0$; e.g., \cite{joh96,joh00}, \cite{see19}, among other literature.}.
For the current computations, we use the solar case as guidance, where the magnetic field strength
at the base of the flux tube is readily assumed as $\eta=B_{0}(z=0~{\rm km})/B_{\rm eq}$ = 0.85
of the equipartition field ($B_{\rm eq}$); see, e.g.,
\cite{faw11} for an analysis on the impact of $\eta$ on LTW simulations for different types of stars.

The equipartition field is the maximum allowed magnetic field strength for quasi-vacant tubes; it is
computed from the horizontal pressure balance equation given by
\begin{equation}
{B^2(z) \over {8\pi}} + P_{i}(z) \ = \ P_{e}(z)
\end{equation}
\citep[e.g.,][]{spr81,her85}.
In order to compute the equipartition field strength $B_{\rm eq}$, we set the internal gas pressure
$P_{i}(z=0~{\rm km})=0$, with
the external gas pressure determined from our initial models; here $P_{e}
(z = 0~{\rm km}) = 1.59 \times 10^{5}$~dyne~cm$^{-2}$ entails
$B_{\rm eq} (z=0~{\rm km}) = 1998$~G.  The magnetic filling factor
$f_{0} (z=0~{\rm km})$ can be computed from the  empirically determined relationship
between $B_{0}f_{0}$ and the rotational period $P_{\rm rot}$ given as
\begin{equation}
{B_{0}f_{0}(z=0~{\rm km})} \ = 238 \ - 5.51 P_{\rm rot} \ .
\end{equation}
The latter relation has been deduced by \cite{cun99} based on data from \cite{rue97}.
Taking $B_{0}$ = 0.85~$B_{\rm eq}$ enables us to compute the value of the magnetic filling factor at the stellar surface.
For the current computations, we consider two values of $P_{\rm rot}$ = 42.2 days and 38.8 days; see
\cite{hen00} and \cite{bou18}, respectively.  The corresponding photospheric magnetic filling factors are given
as $f_{0} = 0.3\%$ and $1.4\%$, respectively.   Figure~2 shows the two constructed magnetic flux tubes based
on these magnetic filling factors.

%
\begin{figure}
\centering
  \includegraphics[width=0.85\linewidth]{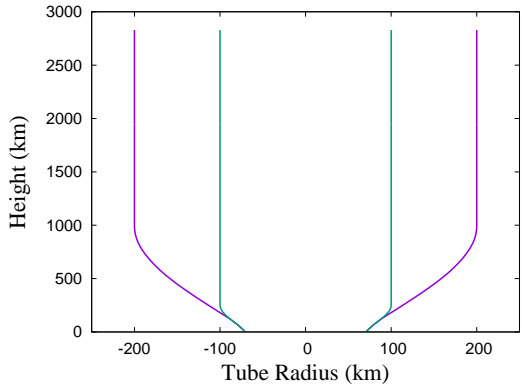}
  \caption{Initial magnetic flux tube models for 55~Cnc.  The photospheric
magnetic filling factor (main models) is given as $f_0$ = 0.3\% (purple).
We also discuss some models with $f_0$ = 1.4\% for comparison (green).
The tube opening radii are given as 100~km and 200~km, respectively.}
\end{figure}


\subsection{Main Radiative Contributors}

The main radiative contributors included in the current computations stem from the H$^{-}$,
Ca~II, and Mg~II ions.  The radiative losses from singly ionized two-level atoms, i.e., the
Ca~II~K and Mg~II~{\it k} lines, are scaled to represent the respective multi-level atoms.

The underlying atomic models entail the five-level-based Ca~II lines (K, H, IRT) and the
three-level-based Mg~II lines ({\it k} and {\it h}), besides the respective continua;
the scaling is done in order to accurately account for the total radiation losses.
The following factors have been adopted: for Mg~II, the factors of 1.50 and 1.53 as
used for monochromatic waves and spectral waves, respectively, and for
Ca~II, the factors of 4.71 and 4.68.
The total radiative damping function (in erg~g$^{-1}$K$^{-1}$s$^{-1}$) is given as
\begin{equation}
\left.{dS\over dt}\right\vert_{\rm Rad} \ = \
\left.{dS\over dt}\right\vert_{{\rm H}^-}
+\left.{dS\over dt}\right\vert_{\rm Ca~II}
+\left.{dS\over dt}\right\vert_{\rm Mg~II} \ .
\end{equation}

Previously, state-of-the-art models, see \cite{ramu03} for the Sun and \cite{faw15}
for other late-type stars, have been developed to derive sets of radiation correction
factors.  These allow the consideration of adequate radiative energy losses given by
multi-level atomic models in the context of time-dependent wave computations.  
Figure~3 shows the radiation damping functions for Ca~II~K and Mg~II~{\it k}
lines as a function of height and for an LTW model with a magnetic filling factor
of $f_{0}=1.4\%$.  Generally, it is found that the formation heights for Ca~II and Mg~II range
between for 700 km and 1800 km, somewhat depending on the model. Radiative energy losses
are most pronounced behind strong shocks owing to the impact
of shock-shock interaction (see Sect. 3.1) and in models with time-dependent hydrogen
ionization omitted (see Sect. 3.2).

In the following, we will mostly study Ca~II, even though
radiative energy losses by Mg~II are relevant as well; see Paper~II for a more extended
analysis.  For example, for the LTW-NTDI model with $f$=1.4\% at an elapsed time of 3914~s,
we find for the ratio (absolute values) of the radiative damping function Mg~II~{\it k} / Ca~II~K
values between $5.4 \times 10^{-3}$ (minimum) and $2.1 \times 10^2$ (maximum),
with an average given as 1.13, depending on the atmospheric height.

%
\begin{figure}
  \includegraphics[width=1.0\linewidth]{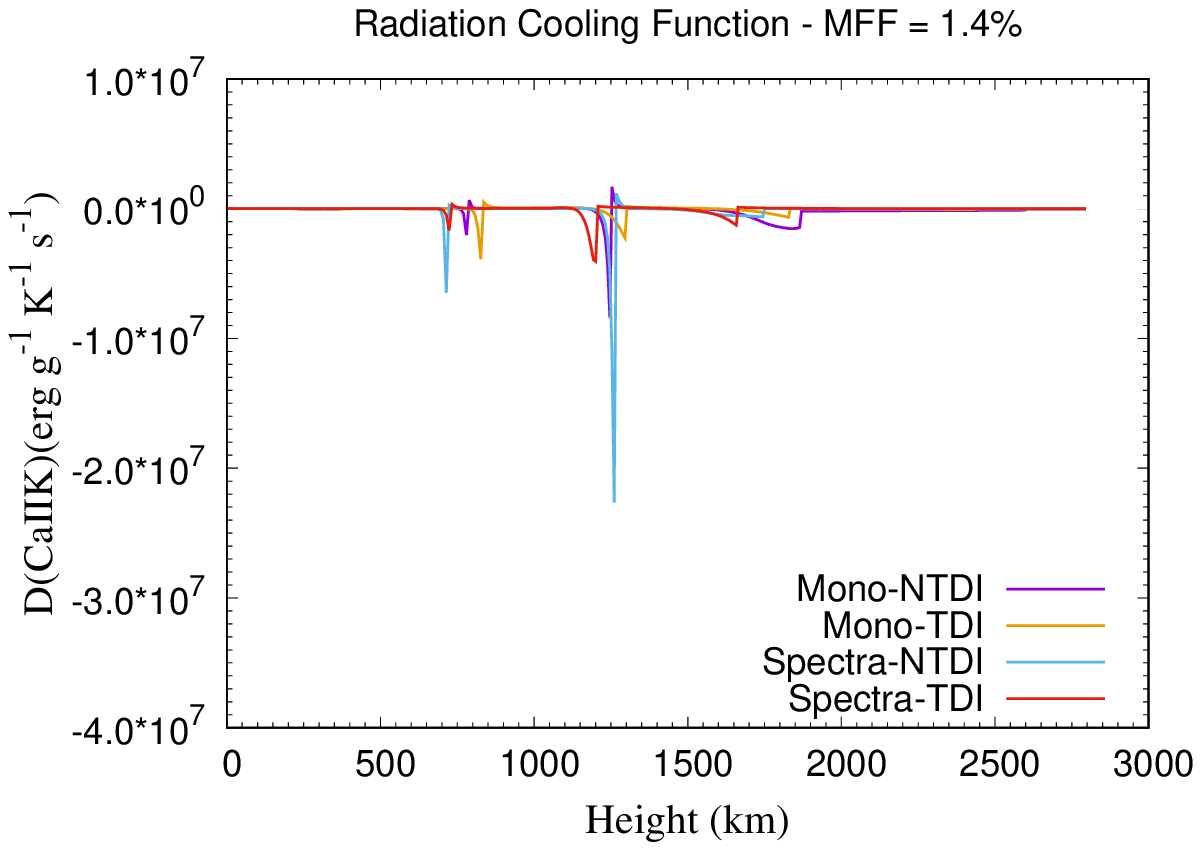}  \\
  \includegraphics[width=1.0\linewidth]{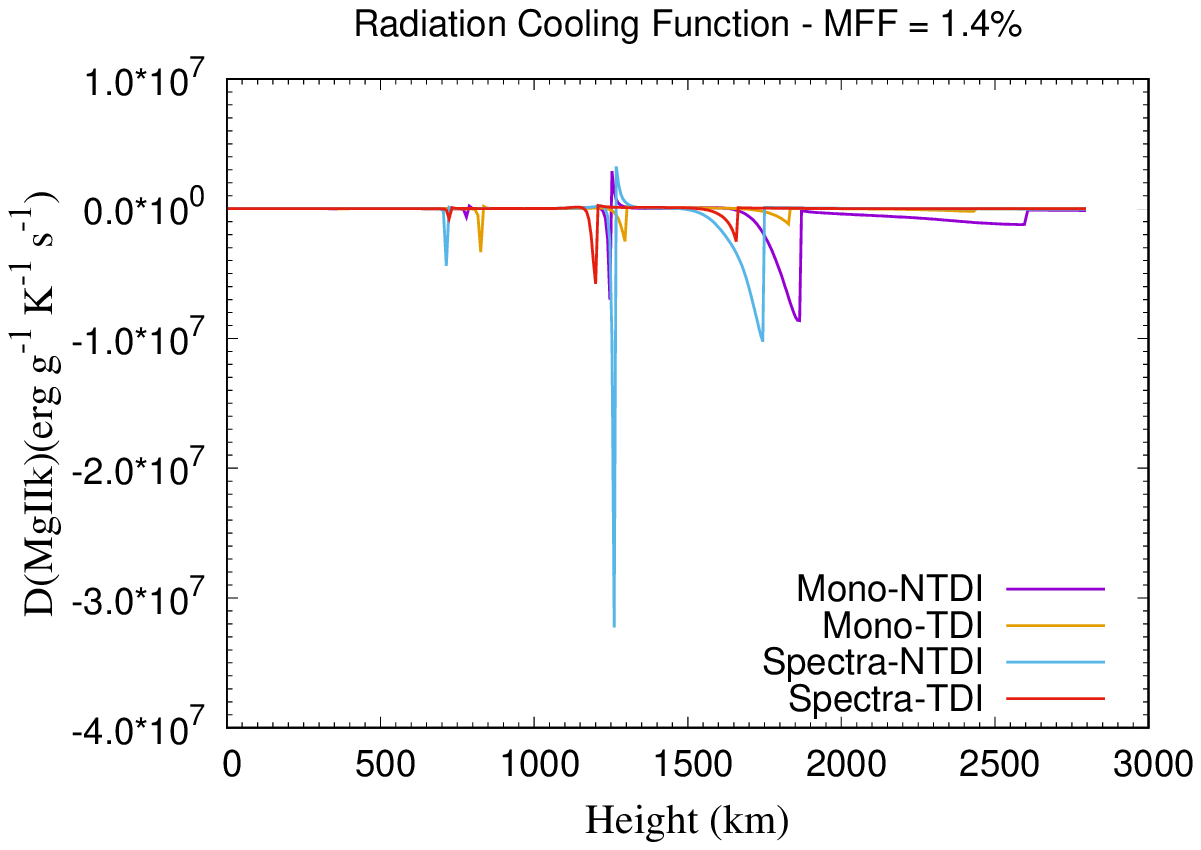}
  \caption{Radiation damping functions regarding the Ca~II~K and Mg~II~{\it k} lines
for different models with $f_{0}=1.4\%$
and an elapsed time of 3940~s.}
\end{figure}


\subsection{Computation of the Emergent Ca~II Emission Fluxes}

A pivotal aspect of time-dependent wave simulations, both pertaining to LTWs and ACWs,
is the calculation of the chromospheric emission based on a detailed implementation for Ca~II.
The chromospheric Ca~II fluxes are computed by considering non-LTE (NLTE) and partial
redistribution (PRD) radiative transfer.  For the adequate representation of the
multi-level atomic model of Ca~II, we rely on previous results by \cite{faw15}.

In this work, the radiation losses by two-level atoms of Ca~II~K line has been scaled up
through the usage of correction factors depending on the type of model, which have been
calculated and tested based on detailed radiative transfer models.  The current computations
are based on multi-ray 1.5-D radiative transfer assuming PRD.
The flux tubes are assumed to be distributed uniformly in a network structure over
the star's surface.  The flux tubes are assumed to be heated by LTWs and the voids
(i.e., non-magnetic regions) by ACWs.  To ensure that the line fluxes
represent the different wave phases, the atmospheres inside the flux tubes are filled
randomly with different wave phases out of total 10 phases. The ACW models are
intended to describe basal chromospheric heating.

To ensure that all surface areas contribute to the computations of the disk-integrated Ca~II~K emission
fluxes, we consider five angles, which are: $\theta = 26^\circ$, $46^\circ$, $60^\circ$, $73^\circ$, and
$84^\circ$.  This approach assures that all surface regions appropriately contribute to the overall Ca~II fluxes.


\section{Results and Discussion}

\subsection{Time-dependent Heating Models}

We obtain time-dependent heating models by starting from the initial atmosphere and inputting
wave energies in form of magnetic and acoustic waves.  They are gradually injected into the atmospheres
over a timespan of about three wave periods.  This process calms down the switch-on effects, thus
allowing to obtain plausible outcomes within a reasonable computational time frame.  Thereafter, waves
are followed in time.  Due to the outward decrease in gas density and pressure, shocks form.
Shock dissipation increases the local temperatures of the chromospheric layers; they also initiate
velocity fields and decisively impact the local and global atmospheric thermodynamics.  The ionization
of the Ca to Ca~II atoms (singly ionized Ca) as well as the Mg to Mg~II (singly ionized Mg)
atoms are local indicators of the increase in temperatures.  After the transmission of (on average)
30 shocks, the atmospheres reach a state of dynamical equilibrium, largely determined by the competition
between the dissipated wave energies and the radiative energy losses.

As part of our study, we consider different types of wave computations,
notably acoustic waves and LTW simulations, regarding magnetic filling factors
of $f_0=0.3\%$ and $1.4\%$, respectively.  In monochromatic wave models,
a dynamic equilibrium is obtained after the original ``switch-on" phase of the
atmosphere subsided.
The initial wave energy flux of the LTWs is about a factor of 5.2 greater
than that of the ACWs; however, owing to the shape of the flux tubes, the
wave energy flux of the LTWs is subjected to a significantly larger degree
of dilution.  As it turns out, there is little difference in the shock
strengths in those models, which is identified as $M_{\rm sh} \approx 1.9$,
largely independent of the atmospheric height.

However, significantly larger shock strength are encountered in wave models
based upon frequency spectra; this is a consequence of shock--shock interaction
also readily identified in previous works \citep[e.g.,][]{cun87,car95,ulm05,faw12}.
This process is especially evident in LTW models --- noting that in some
instances (although only temporarily) shock strengths of
$M_{\rm sh} \gta 20$ are found.  Strong shocks have a profound impact on
dynamic stellar atmospheres, including high temperatures as well as
strong ionization in the post-shock regions (mostly in NTDI models).  Strong temperature
spikes result in strong chromospheric emission, notably in the Ca~II
and Mg~II lines.  However, strong shocks also cause significant events
of momentum transfer, leading to atmospheric expansion, i.e., rapid
episodic outflows, associated with global cooling.  Thus, in those models
complex structures of the overall thermodynamics and atmospheric radiative
environment occur.


\subsection{Effects of Time-dependent Hydrogen Ionization}

The importance of time-dependent ionization of hydrogen (as well as of other species)
in outer stellar layers has been pointed out in many studies; see, e.g., \cite{kne80}
and subsequent work.  These studies indicate that the time scales
of the plasma reaction to, e.g., the passage of shocks and the time scale
associated with hydrogen ionization processes do not coincide.
Sudden fluctuations in the plasma temperature are not mirrored by
instant changes in the hydrogen ionization degree.  The radiative transfer
equations in combination with the statistical rate equations are solved for
the NLTE populations while also taking into account pseudo-partial redistribution
(pseudo-PRD).  The current computations are based on the approach given by
\cite{ramu03} and subsequent work.

Our results show that the time-averaged degree of hydrogen ionizations increases
as a function of height.  Figure 4 depicts a time series of similar wave phases at about
the same time steps.  Here we study LTWs (monochromatic waves) with period of 60~s
and an initial magnetic energy of $F_{\rm LTW}=1.8 \times 10^{8}$  erg~cm$^{-2}$~s$^{-1}$
combined with a magnetic filling factor of $f_0=0.3\%$.  We convey the H, Mg, and Ca
ionization levels as a function of height for models with time-dependent ionization
omitted (see Figs.~4a, 4b, 4c) versus time-dependent ionization included (see Figs.~4d, 4e, 4f).

Comparing Figs. 4a and 4d reveals that the ionization levels of hydrogen are strongly correlated
with the shock strength and the corresponding temperature jumps.  For example,
after the passage at about 2500~s, a steady state of the hydrogen ionization is attained.
However, the saturation level of the degree of hydrogen ionization does not exceed
20\% contrary to the levels of Mg and Ca ionization.  Regarding Figs.~4a and 4c,
a clear saturation in the hydrogen ionization degree is reached around a height of 2300~km.
Some similarities in behavior are also found for models without time-dependent ionization
as in those models strict correlations between, e.g.,
the levels of hydrogen ionization and the instant changes in the local temperatures occur
(see Figs. 4b and 4c).  The hydrogen ionization levels reach a steady state
after around 1250~s, with levels of about 90\% above a height of 2000~km;
this value of hydrogen ionization is considerably higher than in the corresponding TDI models.
In fact, models with TDI experience saturation in regard to hydrogen ionization at mostly 20\%;
see comparison between Figs. 4a and 4d.

In summary, the time-dependent ionization of hydrogen leads to a permanent increase
in Mg~II ionization as a function of height and is uncoupled with the local temperature shocks.
This behavior is not observed in NTDI models, where the degree of hydrogen ionization is
largely coupled with the passage of local shocks and temperature changes.

%
\begin{figure*}
\centering
  \includegraphics[width=0.43\linewidth]{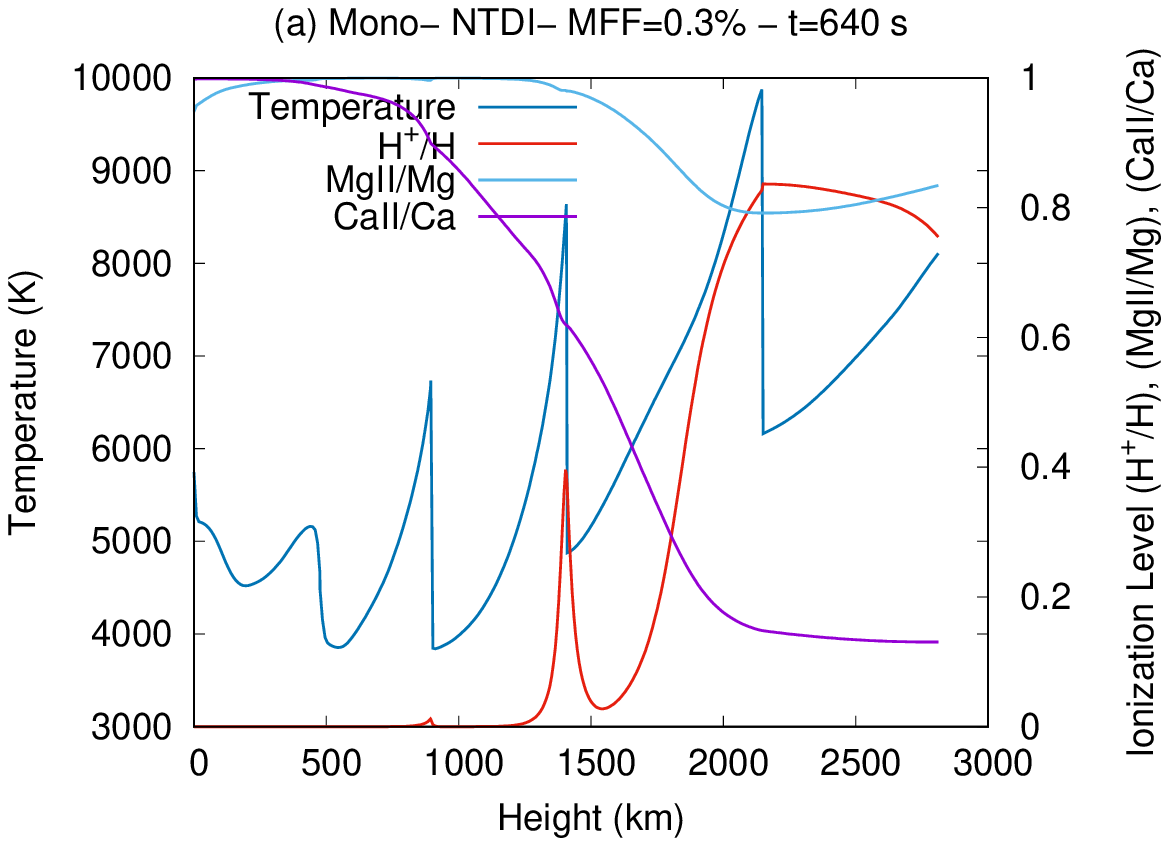} \hspace*{0.5in}
  \includegraphics[width=0.43\linewidth]{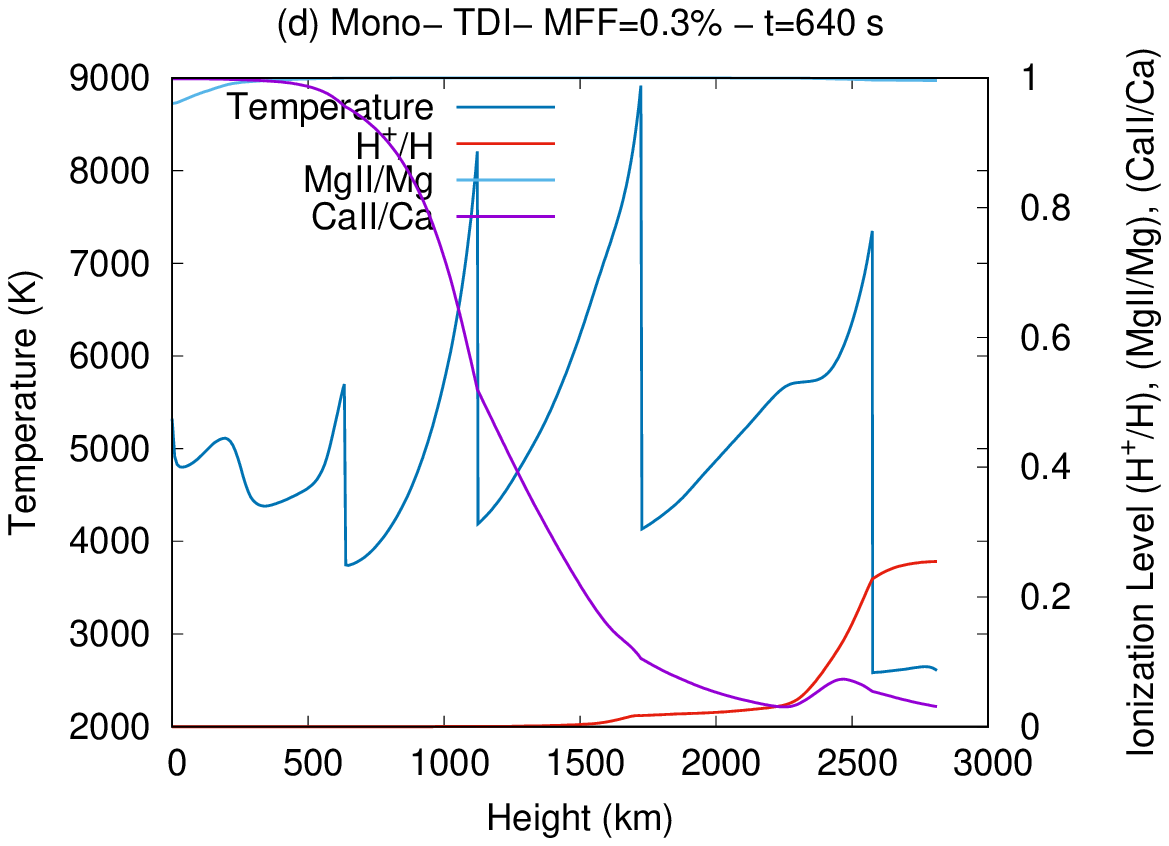}  \\
  \vspace*{0.15in}
  \includegraphics[width=0.43\linewidth]{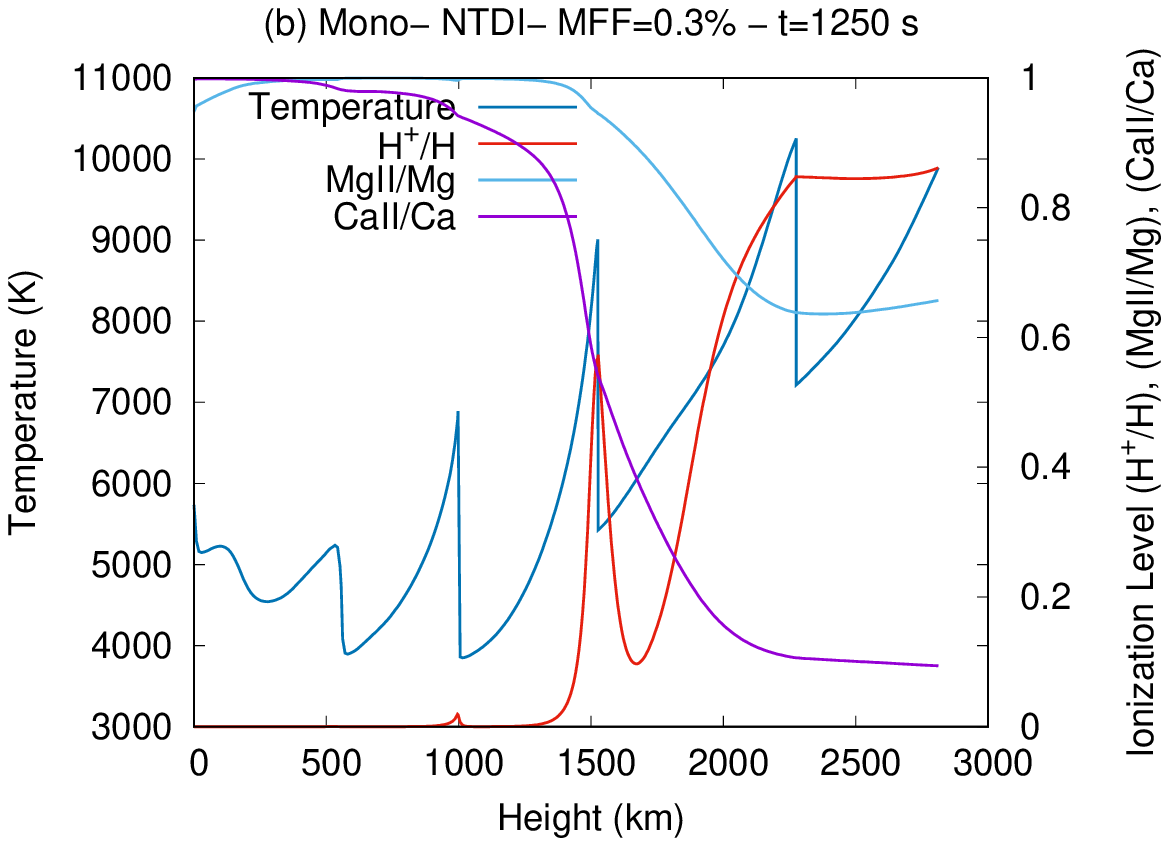} \hspace*{0.5in}
  \includegraphics[width=0.43\linewidth]{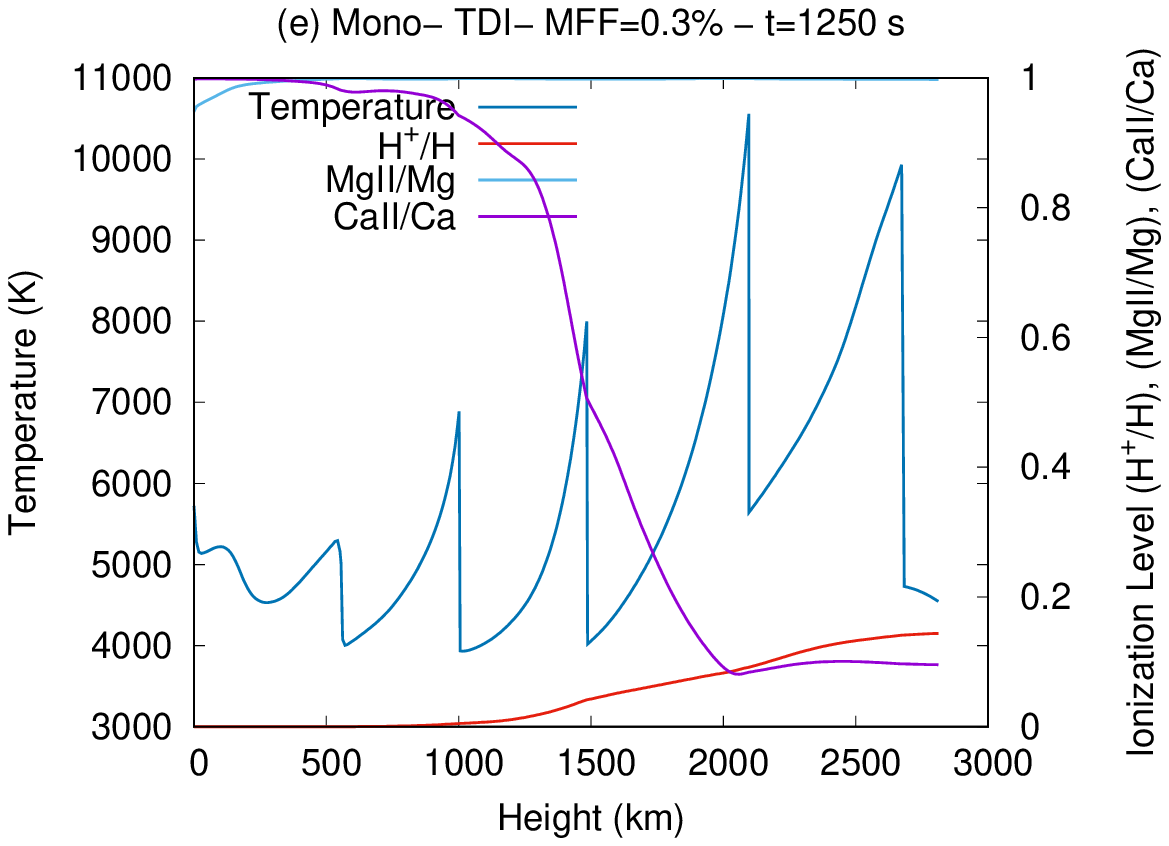}  \\
  \vspace*{0.15in}
  \includegraphics[width=0.43\linewidth]{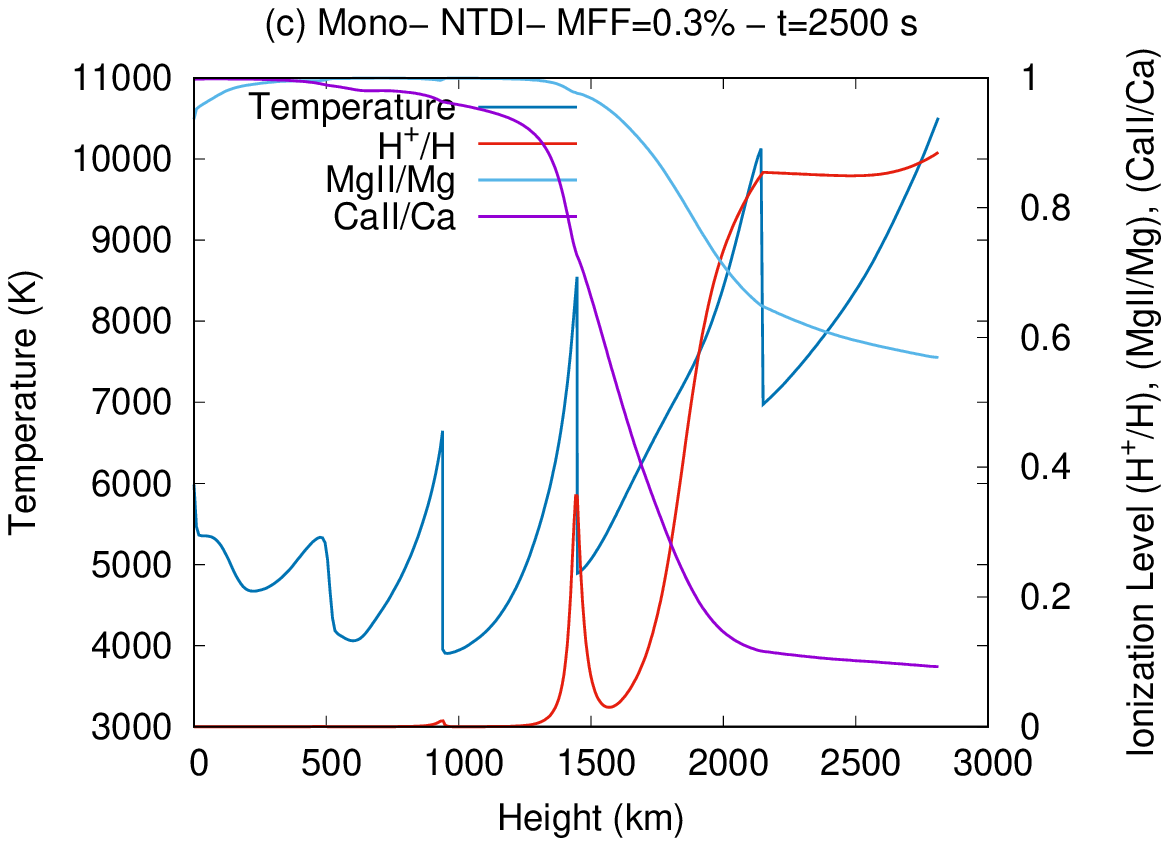} \hspace*{0.5in}
  \includegraphics[width=0.43\linewidth]{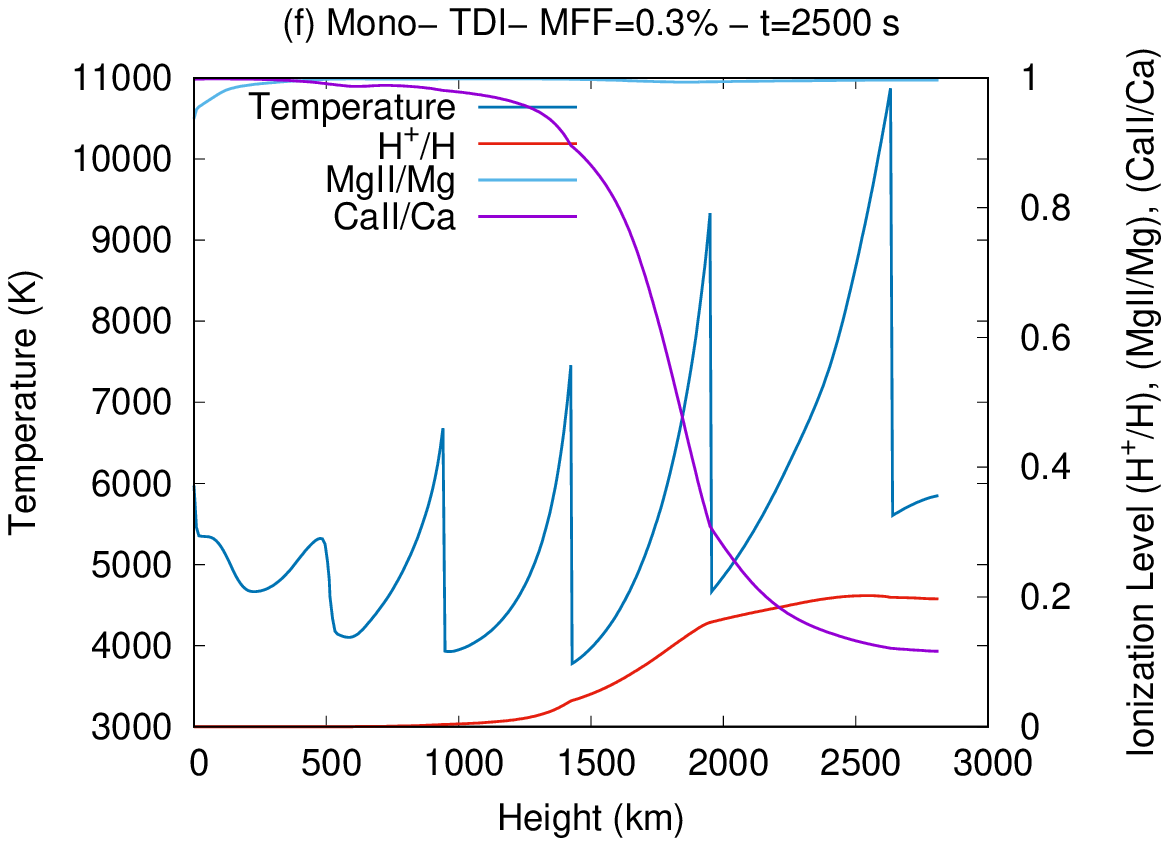}
  \caption{Snapshots of wave phases at different elapsed times for models based on time-independent
  hydrogen ionization (left panel) and time-dependent hydrogen ionization (right panel).  We convey
  the temperature, and the H$^{+}$/H, Ca~II/Ca and Mg~II/Mg ionization levels as a function of the
  atmospheric height.}
\end{figure*}


\subsection{Properties of Time-averaged Atmospheres}

Information on the properties of time-averaged atmospheres is given in
Figures~5 through 8.  In the first two of these figures, we focus on
the temperature, the hydrogen ionization degree, and the Ca~II and Mg~II
ionization of the various models.  For the sets of monochromatic waves,
the averaging in time has been pursued after the ``switch-on" phases
have been subsided, which typically corresponds to about 30 wave periods.
Regarding spectral waves, the ``switch-on" phenomenon persists, as expected;
however, averages have been calculated over sufficiently long time spans
in order to obtain meaningful values.

The mean atmospheric temperatures are found to be highest for monochromatic waves
compared to spectral waves, which is due to the fact that the very strong shocks
initiated by shock--shock interaction (i.e., merging) in spectral wave models
result in strong global cooling, thus counteracting the strong episodic heating
by those shocks.  Furthermore, at large chromospheric heights, the mean
atmospheric temperatures are found to be highest in models of TDI compared to NTDI.
This behavior is attributable to the differences in the shock strengths between
the models.

Moreover, the degree of hydrogen ionization, on average, increases with atmospheric
height in all models.  This behavior is as expected, and it is a direct consequence
of shock heating pertaining to atmospheres that are shaped by a decrease in density
as a function of atmospheric height.  However, the degrees of both Ca~II/Ca and Mg~II/Mg
notably decrease as a function of atmospheric height; this is a direct consequence of the
occurrence of Ca~III and Mg~III due to further ionization as caused by shock heating.
Clearly, the number densities of Ca~III and Mg~III are negligible at lower heights.

Furthermore, models of different magnetic filling factors $f_0$ indicate that larger
filling factors correspond to higher (on average) chromospheric temperatures.  Moreover,
those models are also characterized by the highest Ca~II/Ca ionization rates, especially
in the upper magnetically heated chromospheres.  As previously discussed in the literature
see, e.g., \cite{cun99} and \cite{faw02b}, magnetic wave heating readily exceeds the
heating potential of acoustic waves, especially in models of high magnetic filling
factors.

At large chromospheric heights, there are also notable differences in the
behavior of the mechanical energy fluxes with respect to the models incorporating
TDI or NTDI; see Fig.~7.  TDI models heated by monochromatic or spectral waves
show a more rapid dilution of the wave energy flux as a function of height compared
NTDI models.   This behavior is due to the fact that in TDI models (especially those
based on wave spectra), Ca~II and Mg~II are mostly ionized to Ca~III and Mg~III,
respectively, at those heights.

Another interesting property is that
increases in the initial mechanical energy fluxes (as explored for LTW models) are
essentially inconsequential at intermediate and large chromospheric heights.  This
is a consequence of the well-known limiting shock-strength behavior, as established
for monochromatic waves \cite[e.g.,][]{cun04}, but in terms of the wave energy flux
also encountered by spectral waves.  In fact, as depicted by Table~3, only 1 to 4 \% of the
initial total wave energy flux is converted to Ca~II~K emission, while the lion's share
of that flux (including in models with the initial total wave energy flux increased)
is converted to H$^-$ emission.

Moreover, our simulations also indicate (see Fig. 8) that the mean gas speed increases
as a function of height.  The  highest speeds are obtained in the upper chromosphere
regarding spectral wave models. This outcome is directly associated with the impact
of strong shocks in those models considering that they are able to facilitate the largest
amount of momentum transfer.  The difference between models based on spectral
and monochromatic waves is about a factor of 5.

%
\begin{figure*}
\centering
  \includegraphics[width=0.45\linewidth]{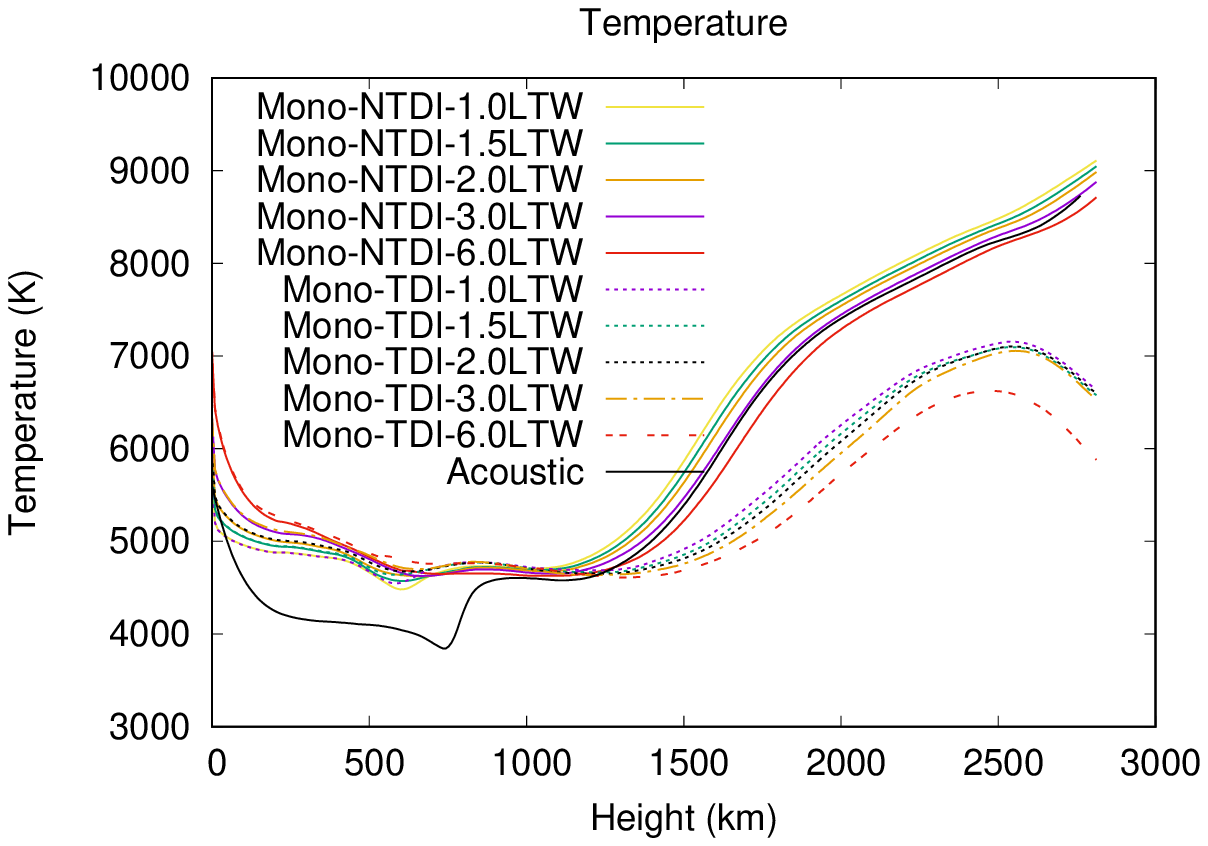}
  \includegraphics[width=0.45\linewidth]{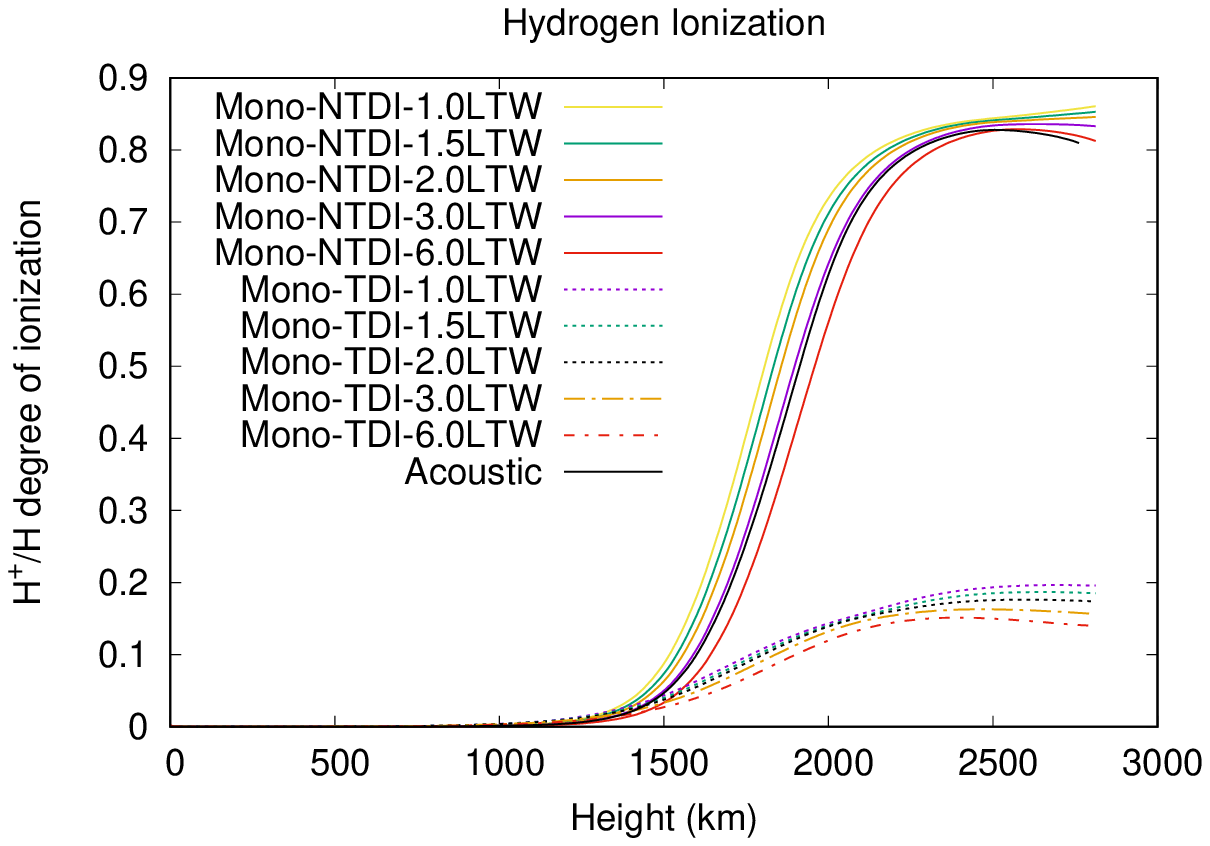}  \\
  \includegraphics[width=0.45\linewidth]{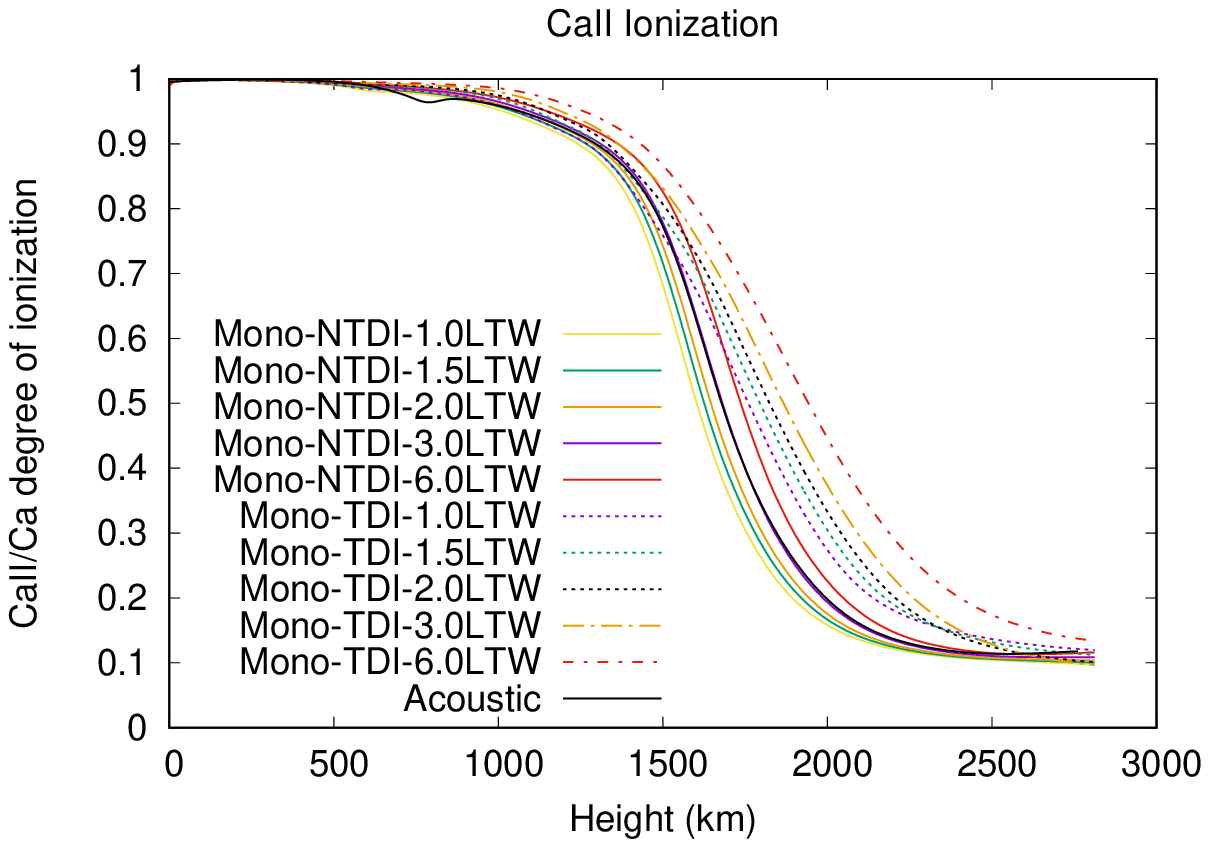}
  \includegraphics[width=0.45\linewidth]{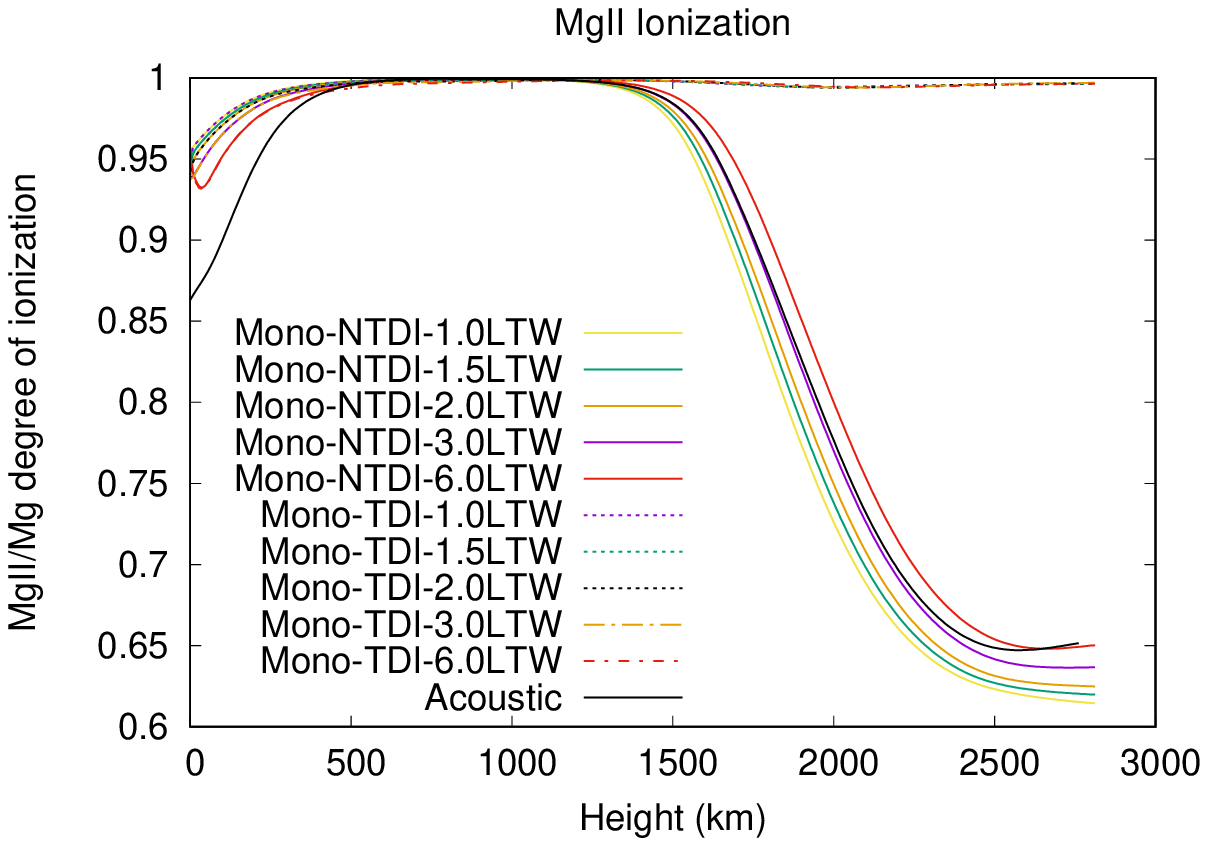}
  \caption{
Results for time-averaged atmospheres for different sets of models based on
NTDI and TDI simulations while assuming different wave energy fluxes and $f_0 = 0.3${\%}.
All models pertain to monochromatic waves.  We depict the atmospheric temperature,
H$^+$/H ionization degree, the Ca~II/Ca ionization degree, and the Mg~II/Mg ionization
degree.  For comparison we also give the results for acoustic waves (monochromatic, TDI).
}
\end{figure*}

%
\begin{figure*}
\centering
  \includegraphics[width=0.45\linewidth]{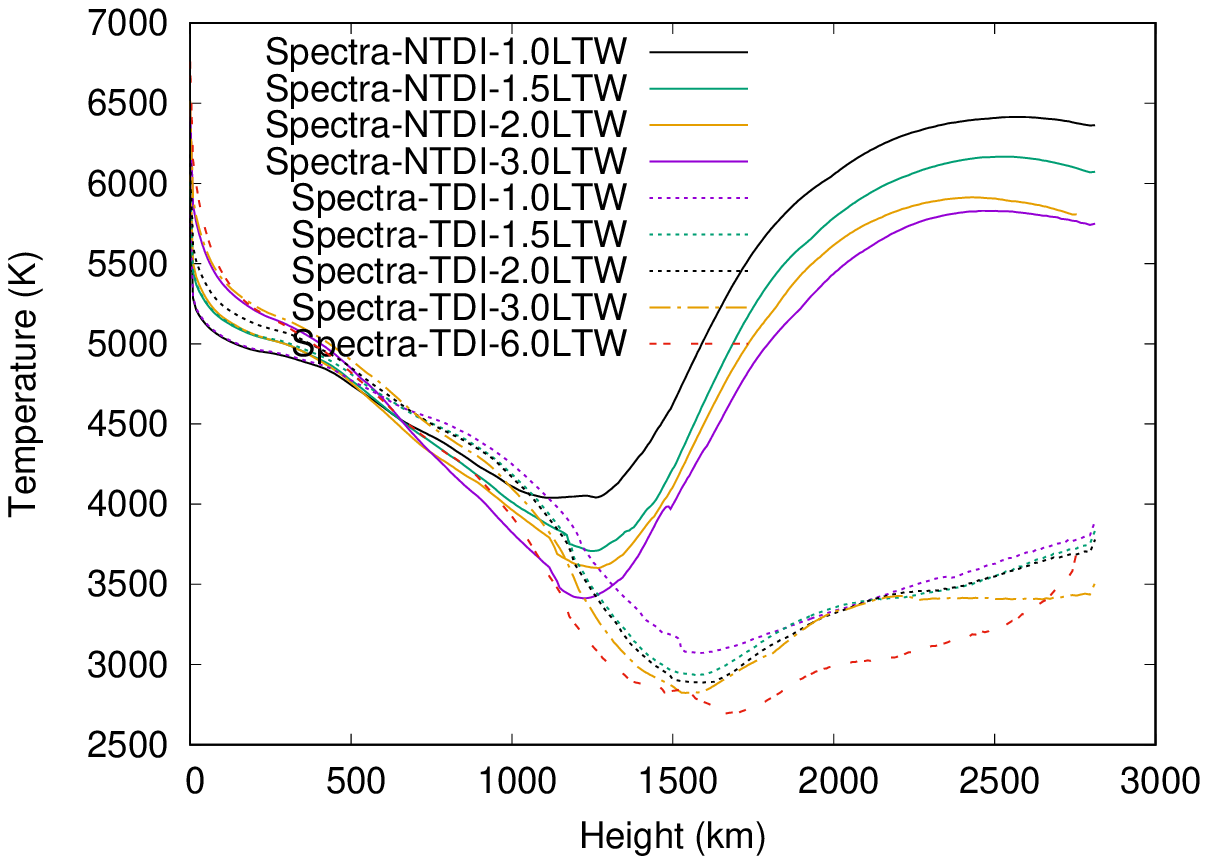}
  \includegraphics[width=0.45\linewidth]{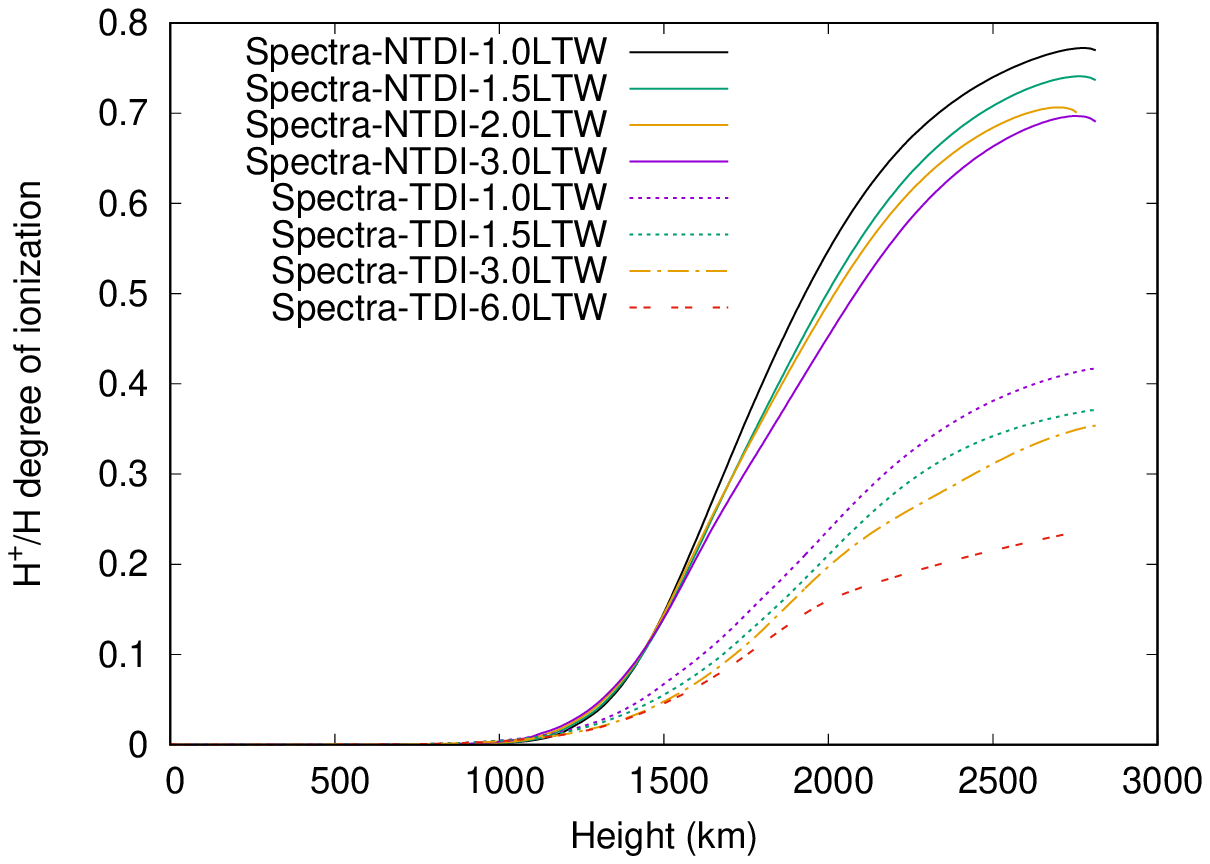}  \\
  \includegraphics[width=0.45\linewidth]{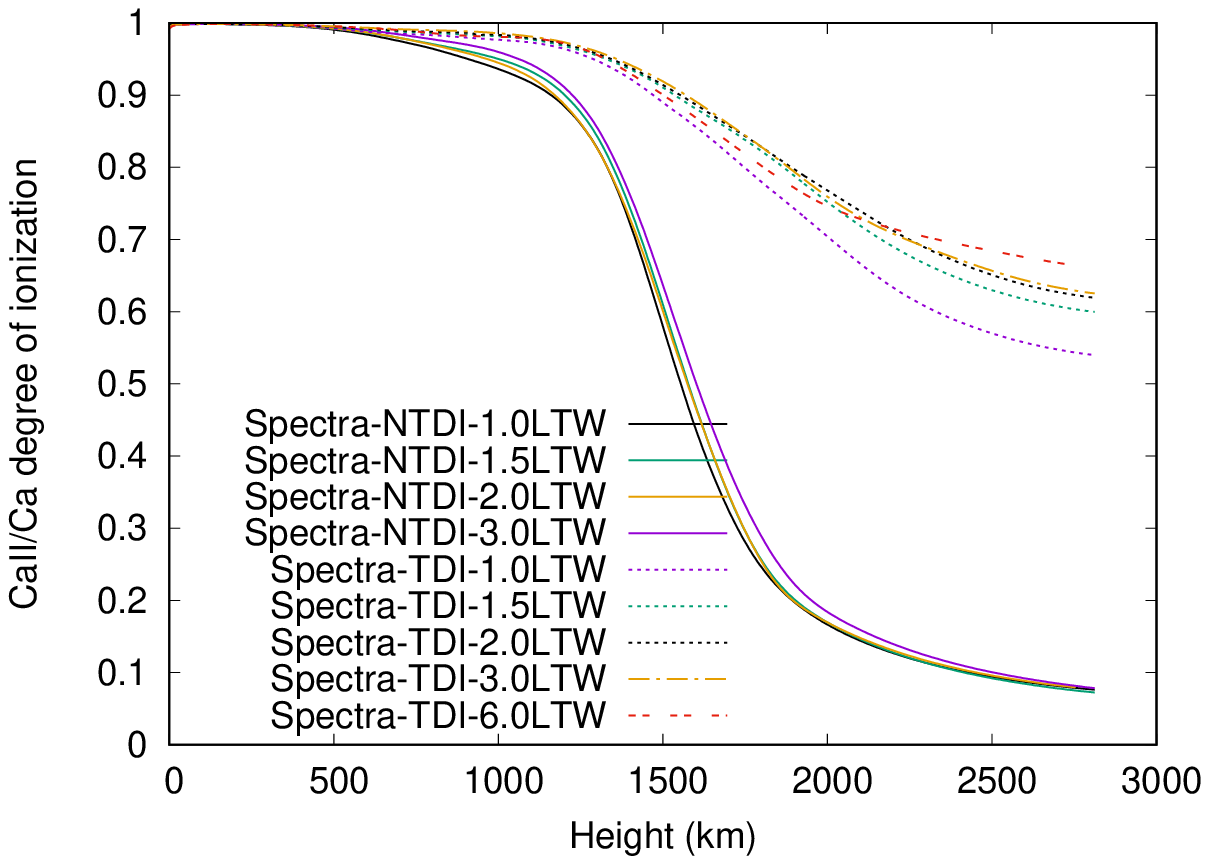}
  \includegraphics[width=0.45\linewidth]{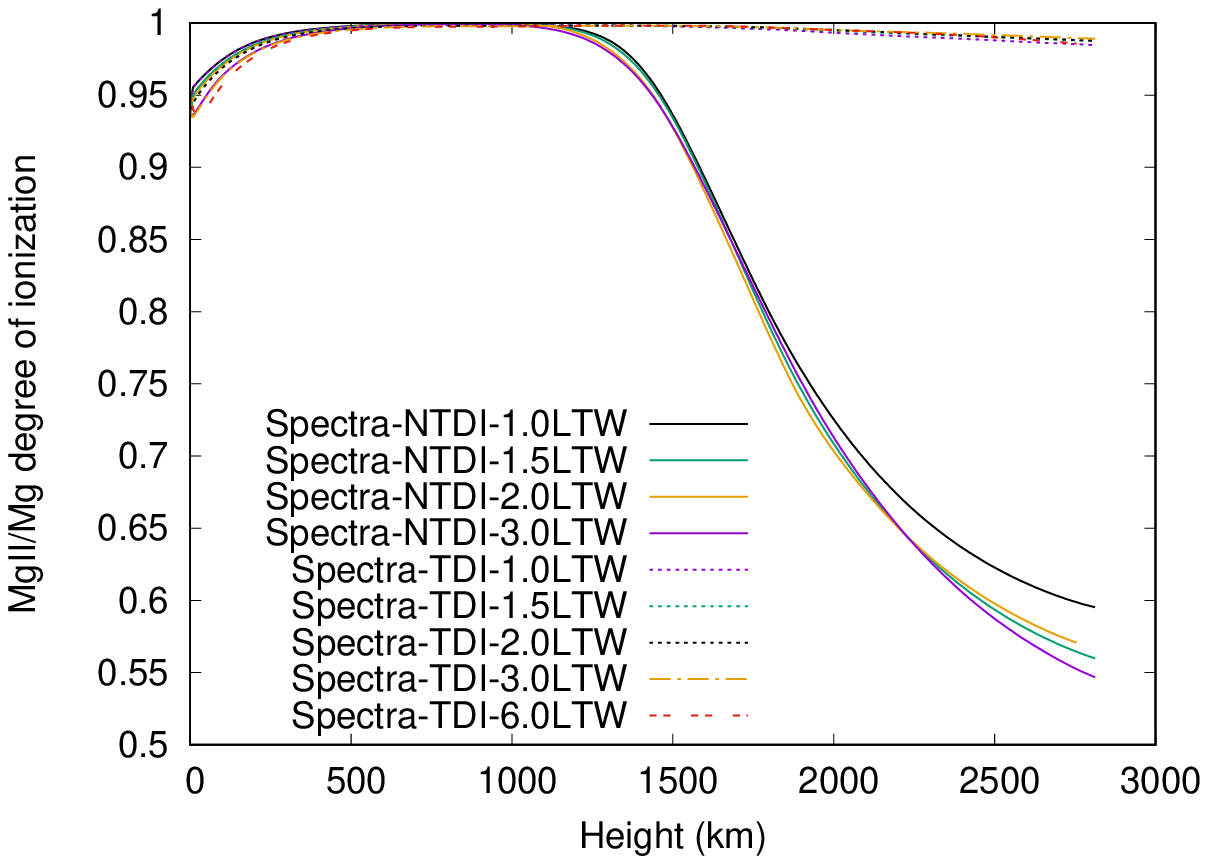}
  \caption{Same as Fig.~5, but pertaining to spectral wave models.}
\end{figure*}

%
\begin{figure*}
  \includegraphics[width=0.5\linewidth]{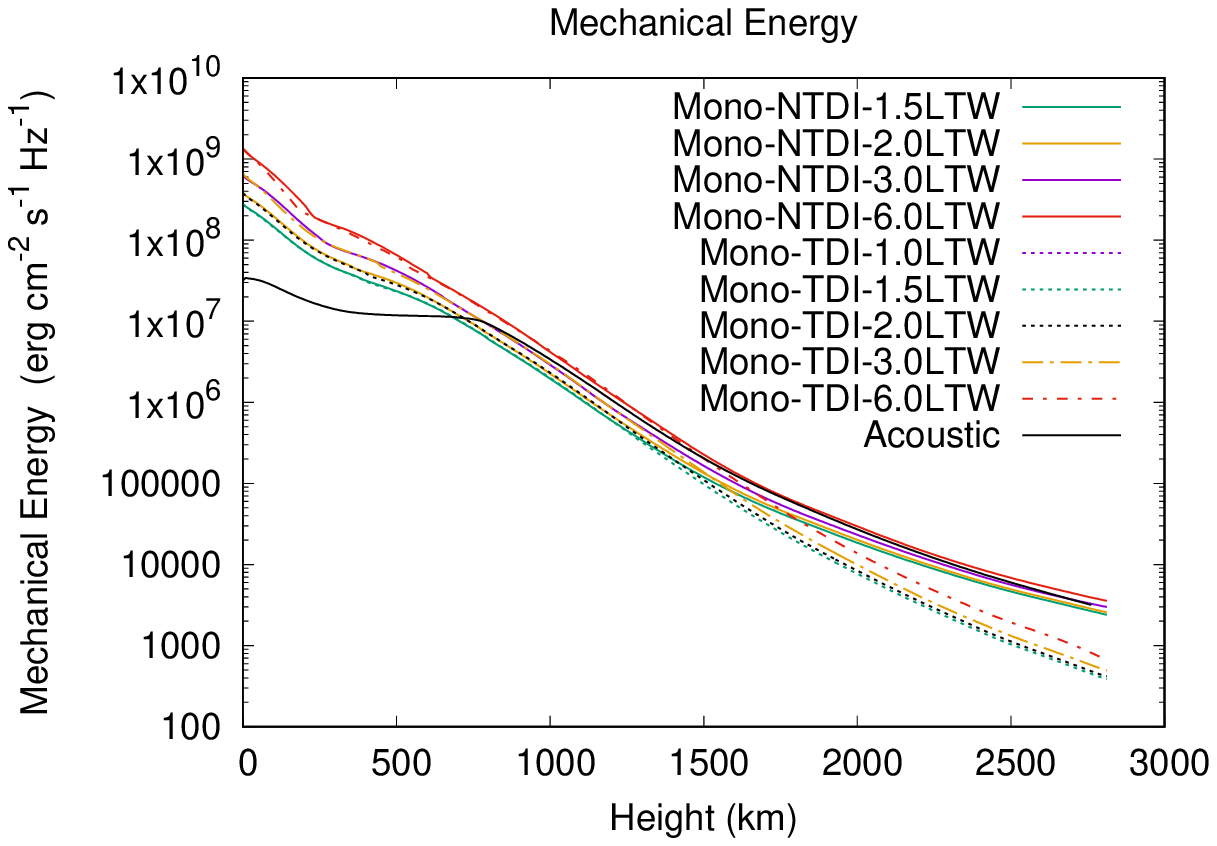}  \\
  \includegraphics[width=0.5\linewidth]{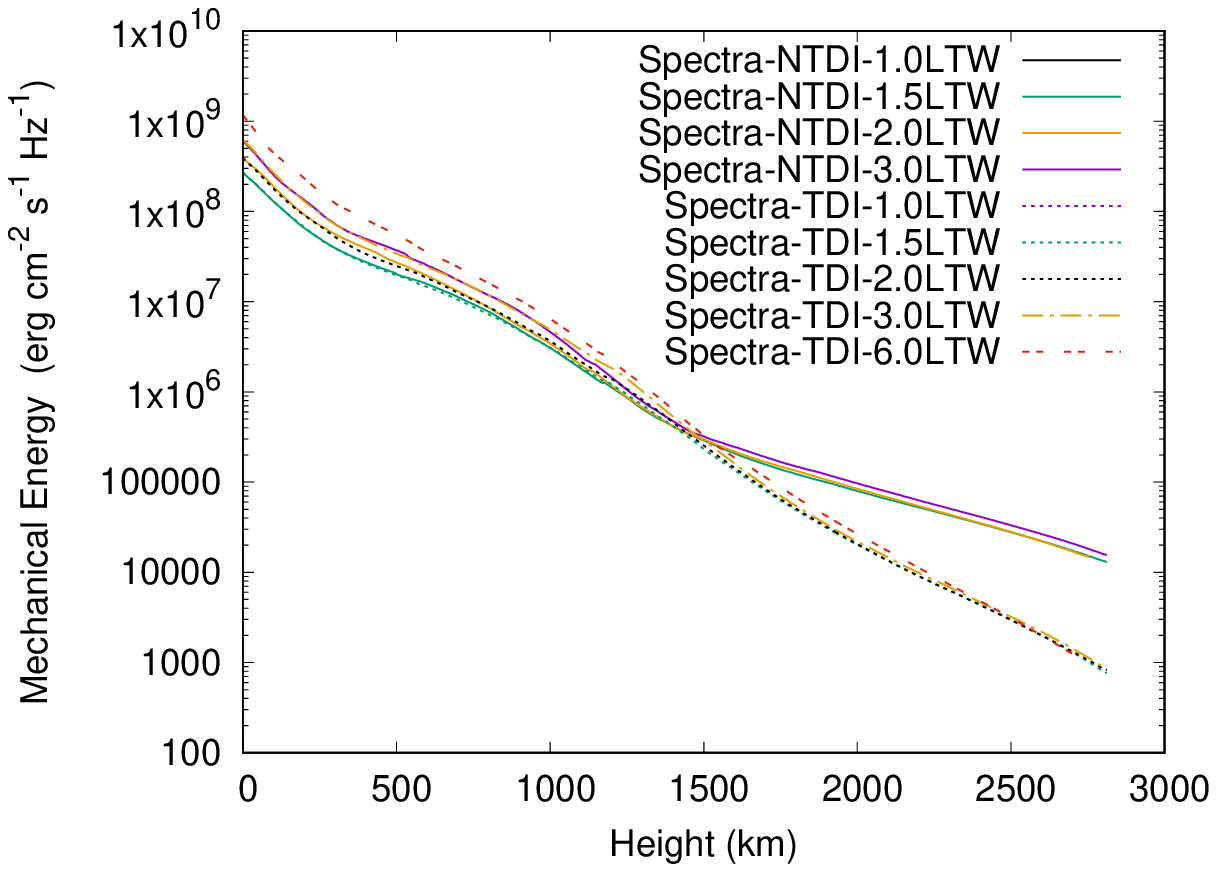}
  \caption{Behavior of the mechanical energy flux as a function of height for
different sets of models.  The top panel conveys monochromatic wave models,
whereas bottom panel conveys spectral wave models.  Here we show both NTDI and
TDI models while also assuming different wave energy fluxes and $f_0 = 0.3${\%}.
For comparison we also give the results for acoustic waves (monochromatic, TDI).}
\end{figure*}

%
\begin{figure*}
  \includegraphics[width=0.55\linewidth]{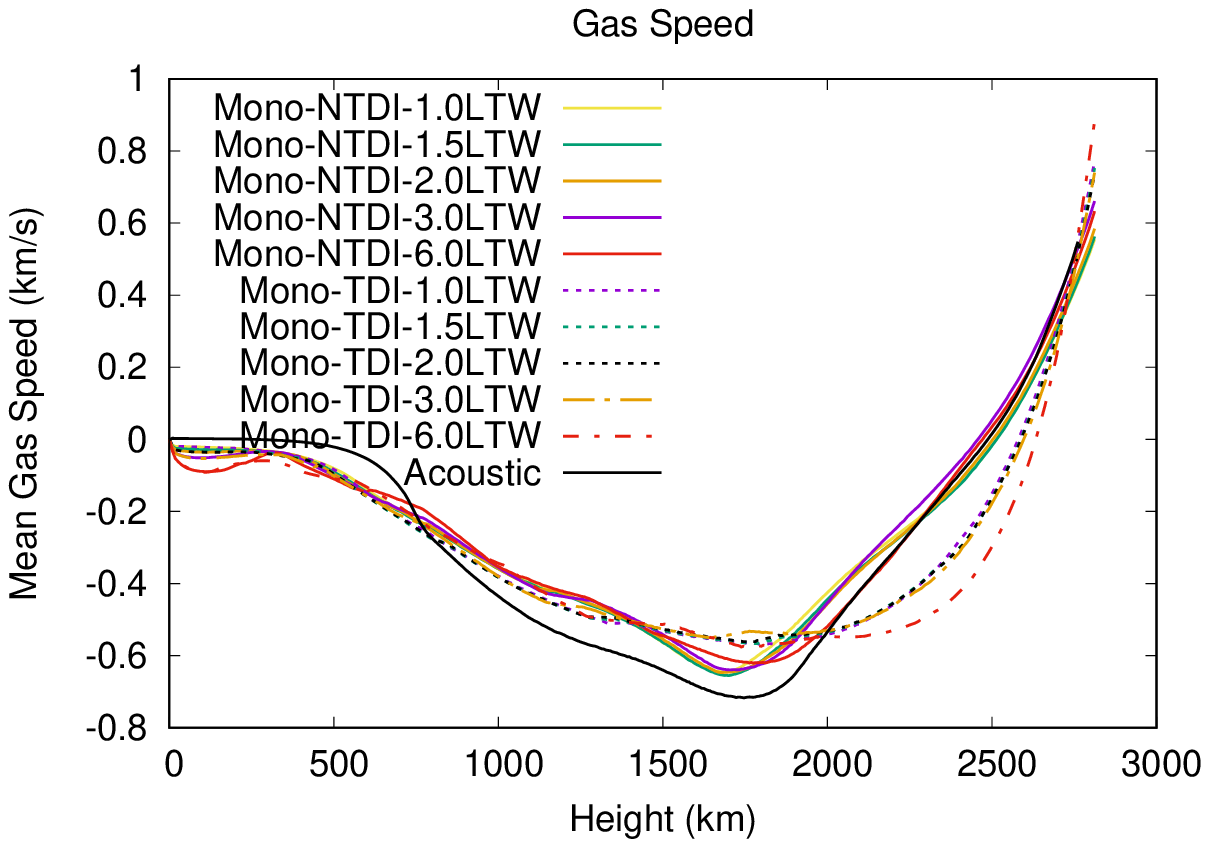}  \\
  \includegraphics[width=0.55\linewidth]{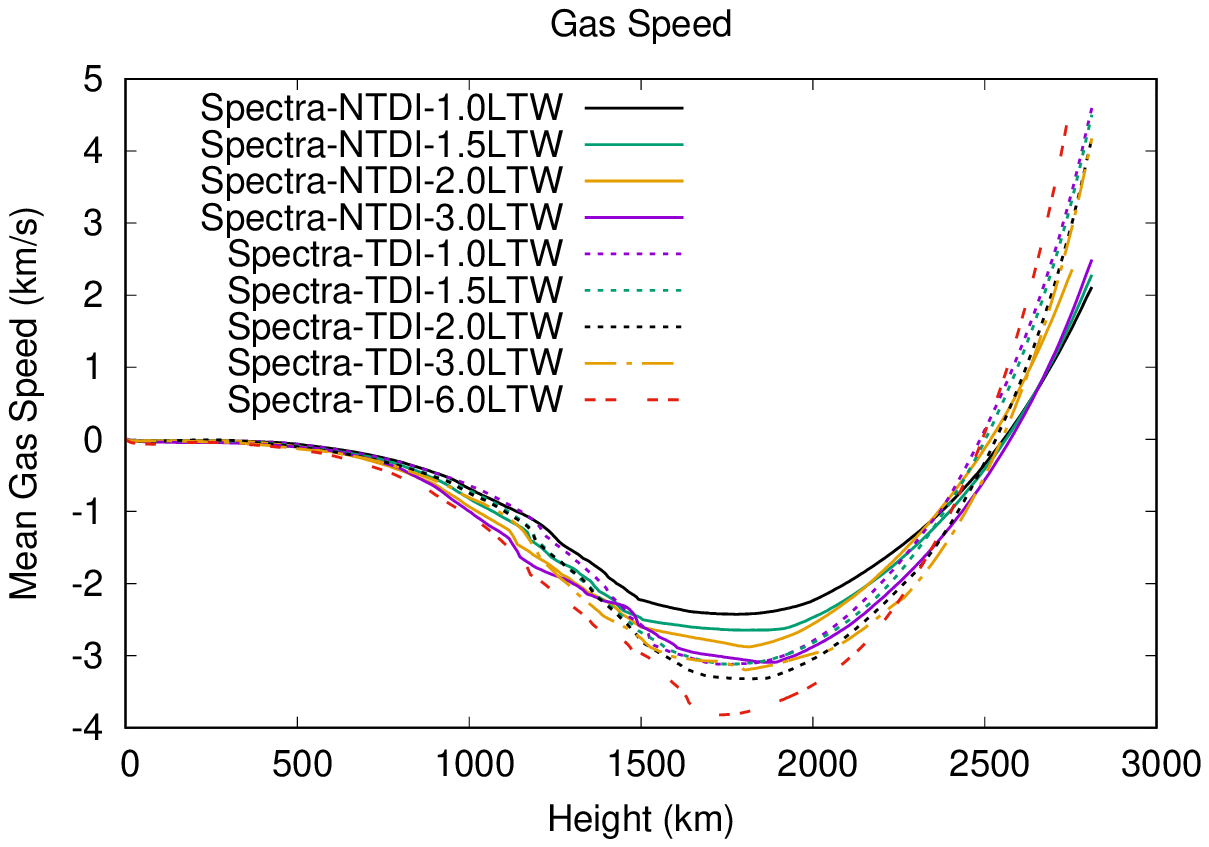}
  \caption{Behavior of the mean gas speed as a function of height for
different sets of models.  The top panel conveys monochromatic wave models,
whereas bottom panel conveys spectral wave models.  Here we show both NTDI and
TDI models while assuming different wave energy fluxes and $f_0 = 0.3${\%}.
For comparison we also give the results for acoustic waves (monochromatic, TDI).
Note the different $y$-scales between the two panels.}
\end{figure*}


\subsection{Studies of Ca~II H+K emission}

\subsubsection{Results for Two-Component Wave Models}

Finally, we focus on studies of Ca~II emission based on our various sets
of models.  The computation of the emergent Ca~II emission fluxes for
the different types of theoretical models follows the approach outlined
in Sect.~2.4 and 2.5.  It ensures that all surface areas contribute to the
computations of the disk-integrated Ca~II emission, which is done through
consideration of different angles of view.

Table~3 conveys the Ca~II~K surface integrated fluxes of the various
models.  Regarding TDI models with initial wave energy fluxes of
1.0 $\times$ LTW   and $f_0 = 0.3${\%} , it is found that the emergent
Ca~II~K for the core emission (defined over a wavelength range of
$\Delta\lambda = \pm 0.3 \r{A}$) is approximately 6\% higher for
spectral waves compared to monochromatic waves --- which is due to
strong shocks occurring in spectral wave models due to shock--shock
interaction, which is largely absent in monochromatic models.  Table~4
shows that this trend also applies to models with $f_0 = 1.4${\%}.  Here the
emergent Ca~II~K for the core emission is about 15\% higher for spectral waves
compared to monochromatic waves.  The corresponding Ca~II~K output for
acoustic waves (spectral models) is reduced by 38\% and 65\%, respectively,
relative to the LTW models with $f_0 = 0.3${\%} and 1.4\%.  Models employing
the NTDI assumption tend to show higher values for the emergent Ca~II~K
core emission as in those models --- at least occasionally --- higher
radiative emissions in the post-shock regions occur.

In our reference models, the computation of the emergent 
Ca~II fluxes are based on instantaneous time-sequences of about 10 wave 
phases with shocks, assessed over timespans of approximately 55~s. In order 
to study the relevance of nonlinearities, as imposed by the shocks,
including the associated effects on the emergent Ca II fluxes, we also
generated time-averages for every five phases over timespans of about 27.5~s.
The comparisons to the instantaneous models reveal that the total emergent
Ca~II fluxes (line cores and wings) are about the same for both cases,
but the time-averaged core flux is reduced by about 20\% compared to the
value obtained by the instantaneous reference model.  Figure 9 shows the
results for an LTW model heated by monochromatic waves with
time-dependent ionization and a magnetic filling factor $f_0=0.3\%$.

%
\begin{figure}
\centering
  \includegraphics[width=0.95\linewidth]{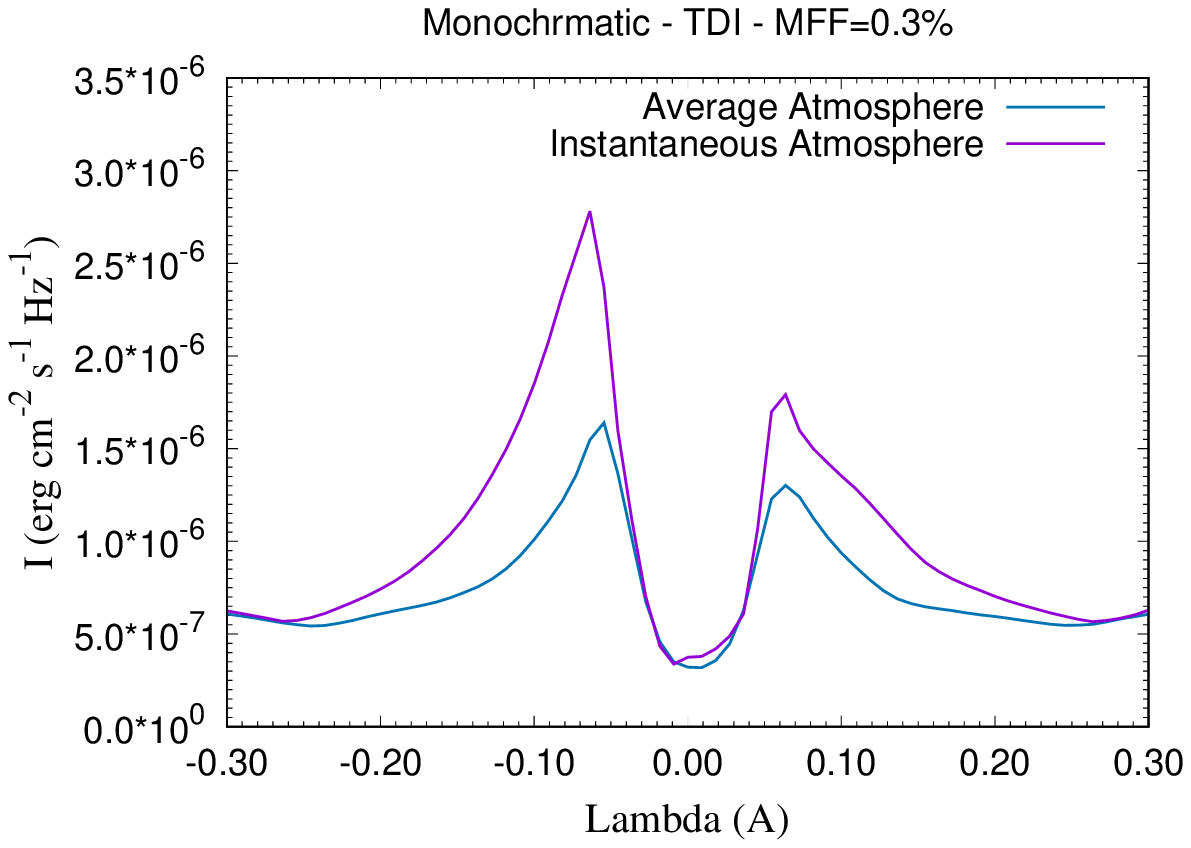}
  \caption{Ca~II~K profiles for an LTW TDI model heated by monochromatic waves
based on the instantaneous and the time-averaged atmosphere.}
\end{figure}


\subsubsection{Consideration of Mode Coupling}

Our main models consider two-component chromospheres based on ACWs and LTWs.
However, we augmented this approach through the consideration of
torsional magnetic tube waves (TTWs); e.g., \cite{nob03} for a detailed analysis of TTW
wave energy generation with applications to the Sun.  TTWs are readily excited in
flux tubes through foot-point motions.  In the context of this work, adopted wave
energy fluxes of TTWs are expressed in multiples of the wave energy fluxes of LTWs.
This approach is also used in assessments of Ca~II emission.

The power spectrum analysis of the vertical and horizontal velocity fluctuations in
the Imaging Magnetograph eXperiment (IMaX) data revealed the existence of the interactions
between longitudinal and transverse wave oscillations in small scale magnetic features
\citep{sta13}.  \cite{has03} previously showed that the nonlinear coupling
of LTWs and TTWs plays an important role in the heating of upper solar chromospheric layers.
The heights of shock formation for both types of waves have been identified as relatively similar,
and after shocks are formed the mode coupling remains highly efficient, thus
effectively contributing to chromospheric heating and emission.

In the view of these findings, we also computed models while employing linear increases
in the input mechanical energy for LTWs with factors ranging from 0.5 to 6 times
of the initial mechanical energy carried by LTWs; see Table~3 for information.  It was
found that the obtained emitted Ca~II fluxes are strongly correlated with the amount
of input mechanical energy up to a value of about 3.5 $\times$ LTW; however, the emitted fluxes
reached distinct levels of saturation (depending on whether the TDI or NTDI assumption
is employed).  This can be explained by the fact that the heating by wave coupling reaches a
limit, implying that other mechanisms would be required to increase the amount of
wave energies any further.  In case of strong shocks, associated with higher
wave energy fluxes, high amounts of momentum transfer occur entailing global cooling,
which tends to counteract the heating initiated by the strong shocks.
Results for the Ca~II~K core fluxes (defined over a wavelength range of
$\Delta\lambda = \pm 0.3 \r{A}$) for different types of models are displayed in Figure~10.
Here $\Delta\lambda$ is calculated relative to the unheated atmosphere; thus, the subtraction
of the photospheric contribution is readily considered.

%
\begin{figure*}
\centering
  \includegraphics[width=0.65\linewidth]{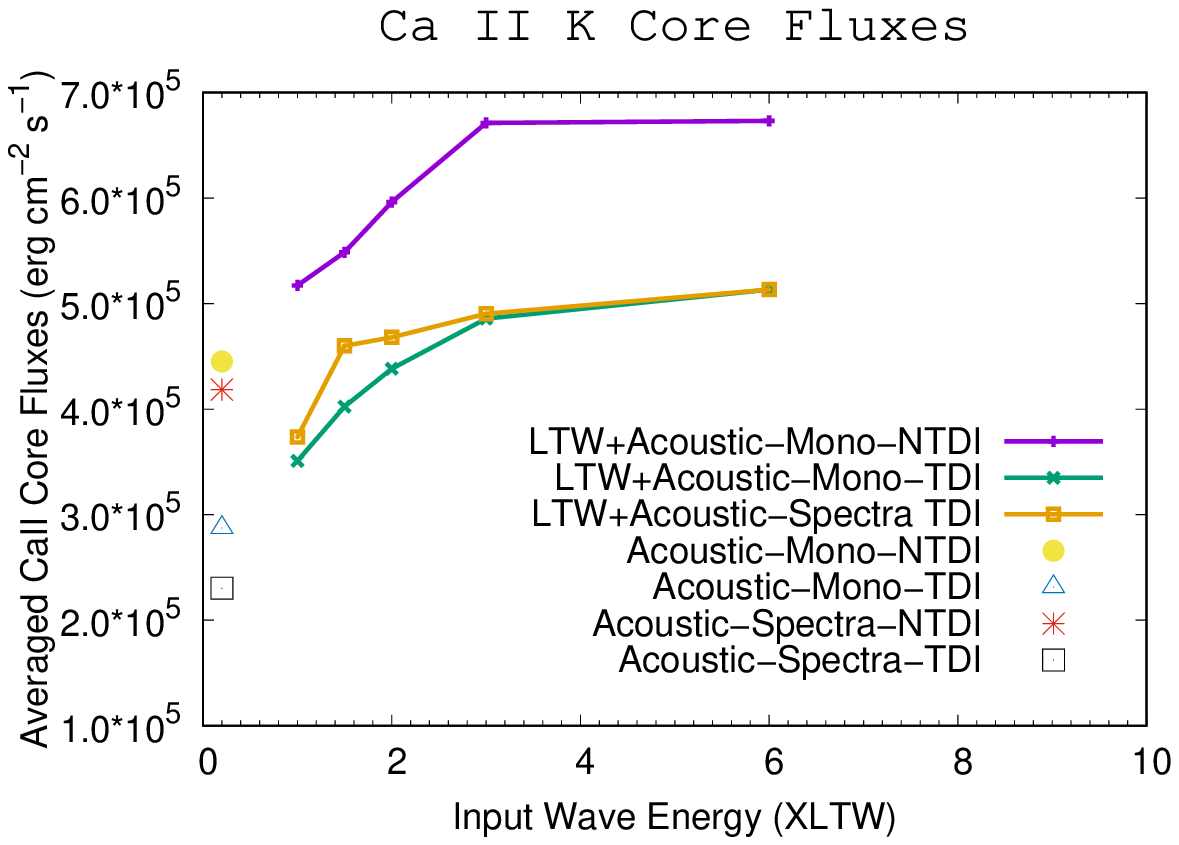}
  \caption{Averaged Ca~II~K core fluxes for different types of models as a function of
multiples of LTW wave energies based on $f_0$ = 0.3\%.
Results are given for monochromatic NTDI and TDI models,
as well as spectral TDI models.  Moreover, we also give the results for the various models
based on acoustic waves.}
\end{figure*}

%
\begin{table*}
	\centering
	\caption{Ca~II~K SURFACE INTEGRATED FLUXES}
\begin{tabular}{llcccc}
\noalign{\smallskip}
\hline
\noalign{\smallskip}
Input Energy & Item & Mono-NTDI & Mono-TDI & Spectra-NTDI & Spectra-TDI \\
\noalign{\smallskip}
\hline
\noalign{\smallskip}
1.0~LTW  &  Total Flux  &  1.43E+7  &  1.40E+7  &  1.61E+7  &  1.62E+7  \\
...      &  Core  Flux  &  5.17E+5  &  3.51E+5  &  4.68E+5  &  3.73E+5  \\
1.5~LTW  &  Total Flux  &  1.52E+7  &  1.41E+7  &  9.33E+6  &  2.11E+7  \\
...      &  Core  Flux  &  5.49E+5  &  4.02E+5  &  4.14E+5  &  4.60E+5  \\
2.0~LTW  &  Total Flux  &  1.75E+7  &  1.46E+7  &  3.30E+7  &  2.35E+7  \\
...      &  Core  Flux  &  5.96E+5  &  4.38E+5  &  4.31E+5  &  4.68E+5  \\
3.0~LTW  &  Total Flux  &  2.82E+7  &  1.95E+7  &  3.32E+7  &  3.64E+7  \\
...      &  Core  Flux  &  6.71E+5  &  4.86E+5  &  5.27E+5  &  4.90E+5  \\
6.0~LTW  &  Total Flux  &  6.07E+7  &  5.61E+7  &  ...      &  5.61E+7  \\
...      &  Core  Flux  &  6.73E+5  &  5.13E+5  &  ...      &  5.13E+5  \\
\noalign{\smallskip}
\hline
\noalign{\smallskip}
\end{tabular}
\end{table*}

%
\begin{table*}
	\centering
	\caption{Ca~II~K SURFACE CORE INTEGRATED FLUXES}
\begin{tabular}{llcccc}
\noalign{\smallskip}
\hline
\noalign{\smallskip}
MFF & Model & Mono-NTDI & Mono-TDI & Spectra-NTDI & Spectra-TDI \\
\noalign{\smallskip}
\hline
\noalign{\smallskip}
0.0  & ACW   & 4.45E+5  & 2.87E+5  & 4.19E+5  & 2.30E+5  \\
0.3  & LTW   & 5.17E+5  & 3.51E+5  & 4.68E+5  & 3.73E+5  \\
1.4  & LTW   & 7.13E+5  & 5.69E+5  & 5.73E+5  & 6.54E+5  \\
\noalign{\smallskip}
\hline
\noalign{\smallskip}
\end{tabular}
\end{table*}


\section{Summary and Conclusions}

We pursued various chromospheric heating calculations for 55~Cnc,
a main-sequence star of spectral-type late-G;
this star generally exhibits a low level of chromospheric activity.
This behavior stems from its old age \citep{mam08,bra11,yee17} and
slow rotation \citep{hen00,bou18}.  Previously
developed codes allow us addressing the different steps of the magneto-acoustic
heating picture\footnote{We contend that there is a large body of literature on that
topic.  The following references are given as examples only; however, they
are most closely associated with the code package utilized by the authors.},
comprised of convective energy generation \citep[e.g.,][]{mus94,mus95,ulm01},
the propagation and dissipation of waves through different layers of
the stellar atmosphere \citep[e.g.,][]{buc98,faw02a}, and the emergence
of radiative energy output \citep[e.g.,][]{cun99,faw02a}.

Besides acoustic waves, our models prominently also take into account longitudinal flux tube
waves to address magnetic heating \citep[e.g.,][]{nar96}.
LTWs are constraint by 55~Cnc's stellar parameters, notably its
rotation period, which in the framework of our model allows us to describe
the photospheric and chromospheric magnetic filling factors; see, e.g.,
\cite{cun99} for a discussion of the inherent features and limitations of this
approach.  Previous models for $\epsilon$~Eri based on this kind of approach
were given by \cite{faw18}.  Moreover, we calculated additional models for
55~Cnc that consider the time-dependent  energy deposition by transverse tube
waves in an approximate manner.

The emergent Ca~II~K fluxes, an important component of the overall
chromospheric energy losses, have been computed in detail (multi-ray treatment) assuming
partial redistribution in combination with time-dependent hydrogen ionization (main models).
Here as well as in our previous papers, these fluxes have been
calculated and interpreted for different levels of stellar magnetic activity,
representative of increased magnetic heating and emission
\citep[e.g.,][and references therein]{cun99,faw02a,faw12}.  Nevertheless, our
main focus relates to the low-activity stage of that star, represented by
a magnetic filling factor of $f_0=0.3$\%; however, additional models
with $f_0=1.4$\% have been calculated as well.

We note that the collisional excitation of the Ca~II upper levels of the
H and K lines is proportional to the local electron density (as well as the local
temperature); see, e.g., \cite{ver81}.  In the chromosphere, H ionization is the
major source of electrons; however, the ionization of H is very different for the
NTDI and TDI models and the temperatures are very different as well.  On the other
hand, the Ca~II emission is not very different for the NTDI and TDI models considering
that the Ca~II H+K lines are formed over a wide range of heights and thus temperatures. 
Nevertheless, the electron densities notably differ between those types of models,
especially behind the shocks.

Based on our models, we obtained the following results:

\smallskip\noindent
(1) Both ACWs and LTWs form shocks in the upper photosphere / lower chromosphere
 In case of monochromatic waves, the height of shock formation for LTWs
(with $f_0=0.3\%$) is about 30~km lower in response to the higher wave energy flux.
Generally, higher shock strengths are attained for spectral waves compared to
monochromatic waves.  This result agrees with previous findings
\citep[e.g.,][]{cun98,cun99,faw18}.  Notable differences between the models
occur at larger heights, mostly due the impact of dilution of the wave energy flux.

\smallskip\noindent
(2) The mean atmospheric temperatures are highest in monochromatic wave models
compared to spectral wave models.  Regarding both monochromatic and spectral models,
NTDI models exhibit a spatial increase in temperature contrary to TDI models.
This behavior is attributable to the differences in the shock strengths between
those models.

\smallskip\noindent
(3)  The formation heights for Ca~II and Mg~II range between for 700 km and 1800 km,
somewhat depending on the model.  Radiative energy losses
are most pronounced behind strong shocks owing to the impact
of shock-shock interaction and in models with time-dependent hydrogen
ionization omitted.  Peaks of Ca~II and Mg~II emission behind
shocks (including those of large strengths) do not occur in TDI models owing to the
difference in the time scales between the ionization processes and shock propagation.

\smallskip\noindent
(4) Models of different magnetic filling factors $f_0$ indicate that a larger filling
factor corresponds to a higher (on average) chromospheric temperatures.  Moreover,
those models are also characterized by the highest Ca~II/Ca ionization rates, especially
in the upper magnetically heated chromospheres.  This result is relevant for stages
of enhanced stellar magnetic activity; see, e.g., \cite{faw02b} for previous results
--- as well as for stars other than 55~Cnc.

\smallskip\noindent
(5)  Increased initial mechanical energy fluxes (as explored for LTW models) are
essentially inconsequential at intermediate and large chromospheric heights for the
atmospheric radiative energy losses, a consequence of the limiting shock-strength behavior,
as established and partially also encountered by spectral waves.  In fact,
only 1 to 4 \% of the initial total wave energy flux is converted to Ca~II~K emission,
while most of that flux (including in models with the initial total wave energy flux increased)
is converted to H$^-$ emission.

\smallskip\noindent
(6) At large chromospheric heights, notable differences exist in the behavior
of the mechanical energy flux for models based on TDI versus those based
on the NTDI approximation.  TDI models heated by monochromatic or spectral waves show
a faster dilution of the wave energy flux at larger heights compared to NTDI models.
This is because Ca~II and Mg~II are mostly ionized to Ca~III and Mg~III, respectively,
at those heights, especially in TDI models employing wave spectra.

\smallskip\noindent
(7) Regarding the averaged Ca~II core fluxes, we also investigated the impact
of increased initial wave energy fluxes.  We found that the emitted Ca~II fluxes
are strongly correlated with the amount of the inputted wave energy up to a value
of about 3.5 $\times$ LTW; however, thereafter a saturation level is reached.
The reason is that heating by wave coupling reached its limit (in part caused
by strong cooling associated with the momentum transfer due to strong shocks).
Hence, other mechanisms would be required to further increase the overall
atmospheric thermal energy budget.

\smallskip\noindent
(8) Our study on 55~Cnc, based on the propagation and dissipation of waves,
constitutes another opportunity to illustrate the significance of nonlinearities.
Assuming LTW TDI models with $f_0=0.3\%$ heated by monochromatic waves
based on either the instantaneous or the time-averaged atmosphere indicates that 
the Ca~II~K core flux is significantly reduced if based on a time-averaged
model atmosphere relative to the (realistic) instantaneous atmosphere.

\medskip

The aim of this study was a detailed assessment of LTW and ACW models for
55~Cnc pertaining to magnetic and nonmagnetic regions, respectively.
The next article of this series (Paper~II) will feature a comparison between
our theoretical models and observations.


\section*{Acknowledgements}

This research is supported in part by the Izmir University of Economics (D. E. F.) and
the University of Texas at Arlington (M. C).  Furthermore, we are grateful to M. Sosebee
(UTA, Dep. of Physics) for his assistance with computer graphics.

\section*{Data availability}

The data underlying this article will be shared on reasonable request to the corresponding author.







\bsp	
\label{lastpage}
\end{document}